\documentclass[conference]{IEEEtran}
\usepackage{graphicx}
\usepackage[breaklinks=true,bookmarks=false]{hyperref}
\usepackage{subfigure}
\usepackage{multirow}
\usepackage[numbers]{natbib}
\usepackage{textcomp}
\usepackage{caption}
\ifCLASSINFOpdf
\else
\fi
\begin{document}
%
% paper title
% can use linebreaks \\ within to get better formatting as desired
\title{OptCon: An Adaptable SLA-Aware Consistency Tuning Framework for Quorum-based Stores}%OptCon: a Flexible Workload and SLA-Aware Framework for Consistency Tuning}

% author names and affiliations
% use a multiple column layout for up to three different
% affiliations
\author{\IEEEauthorblockN{Subhajit Sidhanta\IEEEauthorrefmark{1}, Wojciech Golab\IEEEauthorrefmark{2}, Supratik Mukhopadhyay\IEEEauthorrefmark{1} and Saikat Basu\IEEEauthorrefmark{1}}
\IEEEauthorblockA{\IEEEauthorrefmark{1}Louisiana State University, Baton Rouge, Louisiana, USA, Email: \{ssidha1, supratik, saikat\}@csc.lsu.edu}
\IEEEauthorblockA{\IEEEauthorrefmark{2}University of Waterloo, Waterloo, Ontario, Canada, Email:  wgolab@uwaterloo.ca}}

% conference papers do not typically use \thanks and this command
% is locked out in conference mode. If really needed, such as for
% the acknowledgment of grants, issue a \IEEEoverridecommandlockouts
% after \documentclass

% for over three affiliations, or if they all won't fit within the width
% of the page, use this alternative format:
%
%\author{\IEEEauthorblockN{Michael Shell\IEEEauthorrefmark{1},
%Homer Simpson\IEEEauthorrefmark{2},
%James Kirk\IEEEauthorrefmark{3},
%Montgomery Scott\IEEEauthorrefmark{3} and
%Eldon Tyrell\IEEEauthorrefmark{4}}
%\IEEEauthorblockA{\IEEEauthorrefmark{1}School of Electrical and Computer Engineering\\
%Georgia Institute of Technology,
%Atlanta, Georgia 30332--0250\\ Email: see http://www.michaelshell.org/contact.html}
%\IEEEauthorblockA{\IEEEauthorrefmark{2}Twentieth Century Fox, Springfield, USA\\
%Email: homer@thesimpsons.com}
%\IEEEauthorblockA{\IEEEauthorrefmark{3}Starfleet Academy, San Francisco, California 96678-2391\\
%Telephone: (800) 555--1212, Fax: (888) 555--1212}
%\IEEEauthorblockA{\IEEEauthorrefmark{4}Tyrell Inc., 123 Replicant Street, Los Angeles, California 90210--4321}}

% use for special paper notices
%\IEEEspecialpapernotice{(Invited Paper)}

% make the title area
\maketitle
\thispagestyle{plain}
\pagestyle{plain}

\begin{abstract}
%\boldmath
Users of distributed datastores that employ quorum-based replication are burdened with the choice of a suitable client-centric consistency setting for each storage operation. % that maximizes throughput.
 The above matching choice is difficult to reason about as it requires deliberating about the tradeoff between the latency and staleness, i.e., how stale (old) the result is. %is the version of
% the data item resulting from the given operation with respect to the most recent version of the data item.
 The latency and staleness for a given operation depend on the client-centric consistency setting applied, as well as dynamic parameters such as the current workload and network condition.  %is difficult to reason about as it depends on the use case, as well as the workload, network state, and processing delays at a given time.
  We present OptCon, a novel machine learning-based predictive framework, that can automate the choice of client-centric consistency setting under user-specified latency and staleness thresholds given in the service level agreement (SLA). %Application of consistency levels predicted by OptCon yields maximum throughput for an operation while satisfying the given SLA thresholds for latency and staleness.
  Under a given SLA, OptCon predicts a client-centric consistency setting that is \emph{matching}, i.e., it is weak enough to satisfy the latency threshold, while being strong enough to satisfy the staleness threshold. %satisfies the latency and staleness thresholds given in the SLA.
 While manually tuned consistency settings remain fixed unless explicitly reconfigured, OptCon tunes consistency settings on a per-operation basis with respect to changing workload and network state.
%OptCon predicts the client-side consistency setting that produces the highest throughput under the current workload, network and system state while satisfying the SLA with respect to observed consistency and latency.
%We show experimentally that OptCon provides an intelligent combination of strong consistency in certain feasible scenarios as well as supporting weaker forms of consistency in other cases based on the demand of the situation.
    %Also, we provide a comparative analysis of the performance of the various learning techniques used in implementing OptCon
%        can provide researchers and practitioners with the flexibility of choosing the appropriate learning
%        technique based on the respective application domain and business logic.
 Using decision tree learning, OptCon yields 0.14 cross validation error
 %achieves over 97\% accuracy (given by the area under ROC curve)
  in predicting matching consistency settings % that produce maximum throughput, while satisfying
 under latency and staleness thresholds given in the SLA.
  We demonstrate experimentally that OptCon is at least as effective as any  manually chosen
  consistency settings in adapting to the SLA thresholds for different use cases.
  We also demonstrate that OptCon adapts to variations in workload, whereas a given manually chosen fixed
  consistency setting satisfies the SLA only for a characteristic workload. % only for few workload types.
%With different SLAs, we demonstrate using the RuBBOS benchmark that OptCon is at least as effective as any optimal combination of manually configured fixed consistency settings in achieving the SLA conditions. Under varying benchmark workloads, OptCon automatically adapts the consistency settings producing staleness and latency that matches the SLA demands.
%The abstract goes here.
\end{abstract}
% IEEEtran.cls defaults to using nonbold math in the Abstract.
% This preserves the distinction between vectors and scalars. However,
% if the conference you are submitting to favors bold math in the abstract,
% then you can use LaTeX's standard command \boldmath at the very start
% of the abstract to achieve this. Many IEEE journals/conferences frown on
% math in the abstract anyway.

% no keywords

% For peer review papers, you can put extra information on the cover
% page as needed:
% \ifCLASSOPTIONpeerreview
% \begin{center} \bfseries EDICS Category: 3-BBND \end{center}
% \fi
%
% For peerreview papers, this IEEEtran command inserts a page break and
% creates the second title. It will be ignored for other modes.
\IEEEpeerreviewmaketitle

\section{Introduction}\label{sec:intro}

        %\emph{Client-centric consistency} deals with consistency %across the replicas in a distributed datastore
%        observed from the viewpoint of the client application%\cite{Tanenbaum:2006:DSP:1202502}
%        .
  Many quorum-based distributed  data stores \cite{Lakshman:2010:CDS:1773912.1773922,Meiklejohn:2013:RPD:2505305.2505309,Calder:2011:WAS:2043556.2043571}, allow the developers to explicitly declare the desired \emph{client-centric consistency} setting (i.e., consistency %across the replicas in a distributed datastore
        observed from the viewpoint of the client application) for an operation. Such systems  accept the client-centric consistency settings for an operation in the form of a
        runtime argument, typically referred to as the \emph{consistency level}.
    The performance of the system, with respect to a given operation, is affected by the choice of the
   consistency level applied \cite{Terry:2013:CSL:2517349.2522731}. %, which acts as a tuning knob for the system.
   \par From the viewpoint of the user, the most important performance parameters affected by the consistency
   level applied are the latency and client-observed \emph{consistency
anomalies}, i.e., anomalies in the result of an operation observed from the client application, such as stale reads.
%\cite{DBLP:conf/cloud/GolabRAKWG13}) %(i.e., freshness of the value of a data item returned  by a given operation for an operation o data access),
    %\cite{Tanenbaum:2006:DSP:1202502,Terry:2013:CSL:2517349.2522731}
  %\cite{Terry:2013:CSL:2517349.2522731}. % for a given operation on a quorum-based store.
   Consistency anomalies are measured in terms of the  client-centric \emph{staleness}
  \cite{bayou}, i.e., how stale (old) is the version of
 the data item (observed from the client application) with respect to the most recent version.
  %Weak consistency levels require lesser coordination among replicas, and yield higher
%    staleness and lower latency, whereas strong consistency levels yield lower staleness and higher latency.
     According to the consistency level applied, the system waits for coordination
     among a specific number of replicas containing copies (i.e., versions) of the data item accessed by the given operation \cite{Lakshman:2010:CDS:1773912.1773922}. If the
     system waits for coordination among a smaller number of replicas, the chance of getting a stale result (i.e., an older version) increases.  Also, the latency for
      the given operation depends on the waiting time for the above coordination; hence, in turn, depends on the
       consistency level applied.  For example, a
   weak consistency level for a read operation in Cassandra \cite{Lakshman:2010:CDS:1773912.1773922}, like READ ONE, requires only one of the replicas
   to coordinate successfully, resulting in low latency and high chances of a stale read. % (since it does not wait for
%   all the replicas to coordinate, hence does not ensure that the read returns the the most recent value for the data item). %The write consistency level QUORUM, placed higher than ANY in $C_{list}$, requires a quorum (a group) of the replicas to be written. ALL is the highest read consistency level, that requires every replica node in the cluster  to successfully respond with results.
 %OptCon automates the choice of client-side consistency level to satisfy SLA performance requirements.
 Hence, while choosing the consistency level, developers must consider how this choice
  affects the latency and staleness for a given operation.
  \par
 The chosen consistency level must be \emph{matching} with respect to the latency and staleness thresholds specified in the
 given service level agreement (SLA), i.e., it must be weak enough to satisfy the latency threshold, while being strong enough to satisfy the staleness threshold.
    %Thus, the consistency level for a given operation must be \emph{matching} with respect to a given SLA, i.e., satisfy the consistency (measured in terms of staleness) and latency thresholds given in the SLA.
      %, which specifies that the consistency level
  %for a given application must satisfy the programmatically declared correctness conditions. %, whereas the matching consistency level must satisfy the SLA thresholds of latency and staleness. %A \emph{matching} consistency
%  level in OptCon is the consistency level that produces latency and staleness within specified SLA thresholds.
      Consider a typical use case at Netflix \cite{NetflixBlog} where a user browses recommended movie titles. Such use cases require real-time response \cite{NetflixReal-time}. Hence the SLA
      typically comprises low latency and higher staleness thresholds. For the given SLA, workload, and
      network state, the matching choice is a weak read-write consistency level (like ONE/ANY in Cassandra). If the developer applies stronger consistency level, the resulting high latency might violate the SLA.
% \par
% $C_{Op}^{\mathit{mat}} = $ \[ \Set{C_{op}^j}  {C_{op}^j \in C, \; (L_{op}^j \le L_{op}^{\mathit{SLA}})(S_{op}^j \le S_{op}^{\mathit{SLA}})}{.} \]
  \par With the current
 state-of-the-art \cite{Terry:2013:CSL:2517349.2522731}, the developers have to manually determine a matching consistency level for a given
 operation at development time. Reasoning about the above choice  is difficult
 because of: 1) the large number of possible SLAs, 2) unpredictable factors like changing network state and
 varying workload that impact the latency and staleness, and 3) the absence of a well-formed mathematical relation connecting the above parameters
  \cite{Bailis:2012:PBS:2212351.2212359}. This
 makes automated consistency tuning under latency and staleness thresholds in the SLA a highly desirable feature for quorum-based datastores.  %Pileus and Tuba
% \cite{Terry:2013:CSL:2517349.2522731, Ardekani:2014:SGC:2685048.2685077} are the only systems that come close
%  in providing automated consistency tuning. But these systems perform multiple trials on the system and select
%   the SLA row with minimum resultant utility.   %However, in absence of any existing work that models the effect of the client-side consistency settings on the observed staleness and average latency \cite{Bailis:2012:PBS:2212351.2212359}, that choice is currently arbitrary and unreliable, even with skilled and experienced developers.
      \par We present OptCon \footnote{The project is partially supported by Army Research Office (ARO) under Grant W911\-NF1010495.  Any opinions, findings, and conclusions or recommendations expressed in this material are those of the authors and do not necessarily reflect
the views of the ARO or the United States Government.}, a novel framework that automatically determines
      a matching consistency level %that maximizes the throughput
       for a given operation under a given SLA. %Determining the matching consistency level under a given SLA
%       entails performing constrained optimization on a
%      closed form mathematical model \cite{Bailis:2012:PBS:2212351.2212359}, that captures the impact of the consistency level, current workload, and network state, on the observed latency and staleness.
      Due to the absence of a closed-form
      mathematical model %\cite{Bailis:2012:PBS:2212351.2212359}
       capturing the impact of the consistency level, current workload, and network state, on the observed latency and staleness, % capturing the impact of the model  parameters %, including the client-centric consistency settings,
%       on the observed staleness and latency,
  OptCon applies machine learning \cite{Flach:2012:MLA:2490546} to train a model for \emph{predicting} a
      matching consistency level under the given SLA, workload, and network state. %In contrast with the state-of-the-art
%      automated consistency tuning tools namely, Pileus and Tuba \cite{Terry:2013:CSL:2517349.2522731, Ardekani:2014:SGC:2685048.2685077},
%      OptCon can predict the matching consistency level without actual trials. In the absence of a closed form mathematical model capturing the impact of the client-side consistency settings on the observed staleness and average latency, %\cite{wada:cidr, Bermbach},
%       OptCon applies  machine learning to predict the matching consistency settings.
 For the Netflix use case, taking into account the read-heavy workload, the current network state,
and the SLA thresholds, OptCon predicts a weak consistency level.
 The contributions of this paper are: %summarized as follows.
  \begin{itemize}
  \item We introduce OptCon, a novel machine learning-based framework, that can automatically predict a
   matching consistency level that satisfies the latency and staleness thresholds specified in a given SLA, i.e., the predicted consistency level is weak enough to satisfy the latency
   threshold, and strong enough to satisfy the staleness threshold. Using decision tree learning, OptCon yields a cross validation error of 0.14
 %achieves over 97\% accuracy (given by the area under ROC curve)
  in predicting a matching consistency level %that produce highest throughput
   under the given SLA.
   %\item  Under a given SLA, OptCon predicts the matching consistency level with 0.14 cross validation error using decision tree learning.
   \item  %Experimental results demonstrate that OptCon is at least as effective as carefully chosen manually configured fixed consistency levels in: 1) adapting to the different latency and staleness thresholds in the SLA, and 2) adapting to varying workload.
   Experimental results demonstrate that OptCon is at least as effective as any manually chosen
   consistency level in adapting to the different latency and staleness thresholds in SLAs.
  Furthermore, we also demonstrate experimentally that OptCon surpasses any manually chosen fixed consistency level in adapting to a varying workload,
   i.e., OptCon satisfies the SLA thresholds for variations in the workload, whereas a manually chosen consistency
   level satisfies the SLA for only a subset of workload types.
  \end{itemize}

  \section{Motivation} \label{sec:parameters}
     \subsection{Choice of the SLA Parameters} \label{sec:parameters}
     %Here we provide justification for our choice of latency and staleness as the SLA parameters to be satisfied for a given operation.
      %From the viewpoint of the user, latency and staleness are the most important performance parameters
%      \cite{Terry:2013:CSL:2517349.2522731}.
 Following prior research  by Terry et al. \cite{Terry:2013:CSL:2517349.2522731}, we include latency and staleness in the SLA for operations on quorum-based stores. The choice of consistency level directly affects the
 latency for a  given operation on a quorum-based datastore \cite{Lakshman:2010:CDS:1773912.1773922}.
  While Terry et al. \cite{Terry:2013:CSL:2517349.2522731} use categorical attributes for specifying the desired consistency (such as read-my-write, eventual, etc.) in the SLA, we use a more
  fine-grained SLA, that accepts threshold values for the client-centric staleness \cite{DBLP:conf/cloud/GolabRAKWG13}. % as tolerance bounds for client-observed consistency anomalies with threshold values representing the acceptable bounds for the \cite{DBLP:conf/cloud/GolabRAKWG13}.
 Golab et al. \cite{DBLP:conf/cloud/GolabRAKWG13} %\cite{DBLP:conf/cloud/GolabRAKWG13}
  demonstrate that both the proportion and severity of stale results increases from stronger to weaker consistency levels.
 The use of a client-centric staleness metric %\cite{DBLP:conf/cloud/GolabRAKWG13}
  in the SLA enables the developer to specify, on a per-operation basis, the exact degree of staleness of results, that a given client application can tolerate.
     \par %Golab et al. \cite{DBLP:conf/cloud/GolabRAKWG13} demonstrate that the staleness score (severity) does not show a clear pattern with varying throughput. %, whereas latency exhibits a slight increase.
%varying read/write proportions. %remains approximately steady throughout varying read/write proportions. %Also, throughput stays constant for all possible consistency levels.
       Following Terry et al. \cite{Terry:2013:CSL:2517349.2522731},  we do not include throughput in
       the SLA. But it is still desirable to have throughput \cite{PlanetCassandra} as a secondary optimization
        criterion, once the SLA thresholds are satisfied.
        Depending on the SLA, a set
         of consistency levels can be matching with respect to the latency and staleness thresholds given in the SLA. Consider a real-time application (such as an online shopping cart) that demands moderately low latency and tolerates relatively higher  staleness in the SLA.  In such cases, any weaker consistency level (like ONE or ANY in Cassandra %\cite{Lakshman:2010:CDS:1773912.1773922})
          can yield latency and staleness values within the given SLA thresholds. In such cases of
        multiple matching consistency levels, %(where a group of consistency levels yield observed latency and
%        staleness within the given SLA thresholds \footnote{Consider a real-time application (such as an online shopping cart) that demands moderately low latency and relatively higher  staleness in the SLA.  In such cases, any weaker consistency level (like ONE, ANY or QUORUM) can yield latency and staleness values within the given SLA thresholds.}),
 OptCon chooses the one that maximizes throughput. % (the proportion of staleness increases with the throughput \cite{DBLP:conf/cloud/GolabRAKWG13}).
        Also, we do not consider parameters like keyspace size, replication factor, and certain other server-centric parameters in our model, since our focus is optimizing client-centric performance \cite{DBLP:conf/cloud/GolabRAKWG13}. %, not server-centric performance.
   \subsection{Motivation for Automated Consistency Tuning}\label{sec:autocase}
    %Further, %while tuning the consistency settings OptCon also needs to consider the following factors
   The large number of possible use cases \cite{PlanetCassandra}, each comprising different SLA thresholds for
   latency and staleness, makes manual determination of a matching consistency level a
   complex process. %A matching consistency level must also account for the following additional factors which affect the SLA parameters latency and staleness \cite{Lakshman:2010:CDS:1773912.1773922, Terry:2013:CSL:2517349.2522731, Bailis:2012:PBS:2212351.2212359}, namely the operation type, workload nature, and network state, making the choice further complex.  %Thus manual determination of the matching consistency level for an operation necessitates a thorough understanding of: 1) the performance impact of a specific consistency level  for a given set of operations, 2) the background synchronization tasks, and 3) the relationship of workload characteristics and network  state with the client-side performance.
   The latency and staleness are also affected \cite{Bailis:2012:PBS:2212351.2212359} by the following independent variables (parameters): %While choosing a matching consistency level from client applications under given SLA, the resultant impact on latency and staleness needs to be considered.
    %For a write operation on a key-value store, the staleness \cite{Bailis:2012:PBS:2212351.2212359} and average latency depends on the following controlling parameters:
     1) the packet count parameter, i.e., the number of packets transmitted during a given operation, represents the network state \cite{Lakshman:2010:CDS:1773912.1773922},  and 2) the \emph{read proportion} (i.e., proportion of reads in the workload) and thread count, %\footnote{We consider the initial number of threads created by the client as the thread count parameter. This is because: though we tune consistency on a
%     per-operation basis, in the small delay between operations, the number of threads remains practically unchanged.%Thus, we can assume that the thread count parameter remains practically  unchanged during a client operation.%Hence, the initial number of threads created by the client is considered as the thread count parameter.
%     },
 that represent the workload characteristics. %\cite{Lakshman:2010:CDS:1773912.1773922}.
  Manually reasoning about all of these parameters combined together is difficult, even for a skilled and experienced developer.  Further, unpredictable variations in workload, network congestion, and node failures may render a particular consistency level, manually pre-configured at development time, infeasible  at runtime \cite{183989,Terry:2013:CSL:2517349.2522731}. %Thus, pre-configured or manually tuned consistency settings 1) may not be appropriate for a given operation workload and network state, and 2) may not satisfy the particular latency and staleness requirements demanded by a given use case.

\section{Challenges In Automating Consistency Tuning}\label{sec:challenge}
    %Following the reasoning in Section \ref{sec:parameters}, the automated tuning tool must determine the matching consistency level % that maximises the throughput
%     under the current values of the independent variables (parameters). %While making the manual choice of client-side consistency level, developers need to take into consideration the various knob parameters and the given SLA thresholds \cite{Terry:2013:CSL:2517349.2522731}.
      Currently, there is no closed-form mathematical model \cite{Bailis:2012:PBS:2212351.2212359} that represents the effect of the consistency level on the observed staleness and latency, with respect to the independent variables.
      %The mathematical model of Bailis et al. \cite{Bailis:2012:PBS:2212351.2212359} does not provide a closed form
%       relating latency and staleness to the consistency level.
        Bailis et al. \cite{Bailis:2012:PBS:2212351.2212359} base their work upon the simplifying assumption that writes do not execute concurrently with other operations. Hence, it is not clear from \cite{Bailis:2012:PBS:2212351.2212359} how to compute the latency and staleness from a real workload. %Thus, currently, there does not exist a mathematical model to automate the task of determining the matching consistency level under a given SLA, workload, and network condition.
      Pileus and Tuba
 \cite{Terry:2013:CSL:2517349.2522731, Ardekani:2014:SGC:2685048.2685077} are the only systems that provide fine-grained consistency tuning using SLAs. But, instead of predicting, these systems perform actual trials on the system, and select the consistency level corresponding to the SLA row that produces minimum resultant utility, based on the trial outcomes. The trial-based technique can  produce unreliable results due to the unpredictable parameters like network conditions and workload that affect the observed latency and staleness. Thus, predictions based on the outcomes of the trial phase may be unsuitable in the actual running time of the operation. Sivaramakrishnan et al. \cite{Sivaramakrishnan:2015:DPO:2813885.2737981} use a static analysis approach to determine the weakest consistency level that satisfies the correctness conditions declared in the contract. They do not consider the tradeoff between staleness and latency, and cannot dynamically adapt to varying workload and network state. %Also, it requires the users to have knowledge of a declarative language for specifying the contract, and to accurately specify the correctness rules in the contract. %On the other hand, OptCon is the first work that can predict the matching consistency level that maximizes the throughput under a given SLA on a per-operation basis.  %We present OptCon, a framework that applies machine learning based prediction techniques to automate the choice of the matching consistency level for a given operation under a given SLA.
 \def\tuple#1{\langle #1\rangle}
\section{Design of OptCon}\label{sec:design}
% OptCon is the first work that provides automatic tuning of client-side consistency settings on a per-operation basis, using prediction rather than actual trials.
 %Here we give the design of the OptCon framework.
  %Section \ref{sec:designover} presents the overview of the design. Section \ref{sec:arch} presents the architecture of OptCon. Section \ref{sec:logdes} presents the design of the Logger module, and the Section \ref{sec:approach} presents the design of the Learner module.
  \subsection{Design Overview}\label{sec:designover}
   In the absence of a closed-form mathematical model relating the consistency level to the observed latency and staleness, OptCon leverages machine learning-based prediction techniques \cite{Flach:2012:MLA:2490546}, following the footsteps of prior research  \cite{lacurts2014cicada,Winstein:2013:TEM:2486001.2486020,Ravindranath:2013:TCU:2517349.2522717}%\cite{lacurts2014cicada,Ravindranath:2013:TCU:2517349.2522717}
. OptCon is the first work that trains on historic data, and learns a model \emph{$\mathcal{M}$} relating the consistency level and the input parameters (i.e., the workload and network state) to the observed latency and staleness. The model $\mathcal{M}$ can predict a matching consistency level %workload, use case, and network state,
   with respect to the given SLA thresholds for latency and staleness, under the current workload and network
   state. A matching consistency level is weak enough to satisfy the latency threshold, and simultaneously strong enough to satisfy the staleness threshold.
     Our definition of matching consistency level is a modified version of the \emph{correct} consistency level in QUELA \cite{Sivaramakrishnan:2015:DPO:2813885.2737981}. For  multiple consistency levels satisfying the given SLA thresholds, OptCon predicts the consistency level that maximizes the throughput.
      \par In contrast to program verification-based static (compile time) approaches \cite{Sivaramakrishnan:2015:DPO:2813885.2737981},
      OptCon provides on the fly predictions based on dynamic (runtime) values of the input parameters, i.e., the current workload and the network state (Section \ref{sec:autocase}).
       Thus, OptCon tunes the datastore according to any dynamic variations in the workload and any given SLA.
       Furthermore, unlike the
      state-of-the-art trial-based techniques that base decisions on the values of the parameters obtained during the trial phase \cite{Terry:2013:CSL:2517349.2522731, Ardekani:2014:SGC:2685048.2685077},
      OptCon provides reliable predictions (see Section \ref{sec:vary} and \ref{sec:user}), taking into consideration the actual runtime values of the input parameters.
  \begin{table}[!htb]
\centering
\captionsetup{justification=centering}
%\noindent
\scalebox{1.2}{
\begin{tabular}[b]{|c|c|} % centered columns (4 columns)
\hline\hline %inserts double horizontal lines
 \bf Average Latency & \bf Staleness \\ % inserts table heading
 \hline\hline
 $\leq$100ms & $\leq$5ms \\ % inserting body of the table
 $\leq$50ms & $\leq$10ms \\
 $\leq$25ms & $\leq$15ms \\ % [1ex] adds vertical space
\hline %inserts single line
\end{tabular}}
\caption{Example subSLAs}\label{table:sla}
\end{table}
\par Like Terry et al. \cite{Terry:2013:CSL:2517349.2522731}, users provide latency and staleness thresholds
 to OptCon in form of an SLA, illustrated in Table \ref{table:sla}. Each SLA consists of rows of \emph{subSLAs}, where each subSLA row, in turn, comprises two columns containing the thresholds for latency and staleness, respectively. Instead of
 categorical attributes %(such as eventual, read-my-writes, etc.)
  \cite{Terry:2013:CSL:2517349.2522731}, OptCon uses threshold values for staleness \cite{DBLP:conf/cloud/GolabRAKWG13} in the subSLAs.
% Terry et al. uses SLAs structured as an ordered sequence of subSLA rows, ordered by the utility column. On the other hand, subSLAs
% in OptCon are not necessary ordered.
 Unlike Terry et al., subSLAs in OptCon do not have a utility column. Also, subSLAs are not necessary ordered. If the thresholds for a particular subSLA are violated, OptCon marks the operation as failed with respect to the given subSLA. The handling of such failure cases will depend on the application logic, and is not part of the scope of this paper. % Depending on the use case, the operation is retried against any of the next subSLA rows.
\par Instead of modifying the source code of distributed datastores, OptCon is designed as a wrapper over existing distributed storage systems. %It accepts bounds to latency and staleness (i.e., observed consistency) in subSLA format.
 We have implemented and tested OptCon on Cassandra, %\cite{Lakshman:2010:CDS:1773912.1773922},
  which follows the Dynamo \cite{DeCandia:2007:DAH:1323293.1294281} model. %But OptCon can be easily applied to any Dynamo-style systems, like  Voldemort \cite{Sumbaly_servinglarge-scale} and Riak \cite{Meiklejohn:2013:RPD:2505305.2505309}.
 Among the Dynamo-style systems, while Cassandra organises the data in the form of column families, Voldemort \cite{Sumbaly_servinglarge-scale} and Riak \cite{Meiklejohn:2013:RPD:2505305.2505309} are strictly key-value stores.
   Also, Cassandra and Riak  are designed to enable faster writes, whereas Voldemort enables faster reads. %\cite{Sumbaly_servinglarge-scale, Meiklejohn:2013:RPD:2505305.2505309}. %OptCon can be potentially integrated with any such Dynamo-style system with minor % modifications in the wrapper code.
  But the principles governing the internal synchronization mechanisms are similar for all these systems. %\cite{Lakshman:2010:CDS:1773912.1773922, Sumbaly_servinglarge-scale, Meiklejohn:2013:RPD:2505305.2505309}.
   Hence the consistency level choices affect the observed latency and staleness in a similar fashion for all these systems. Thus, OptCon
  %though we have tested OptCon on Cassandra only, it
  can potentially be integrated with any Dynamo-style system. % with minor modifications.
\subsection{Architecture}\label{sec:arch}
\begin{figure}[!htbp]
        \centering
        \includegraphics[width=2in,height=0.8in]
                    {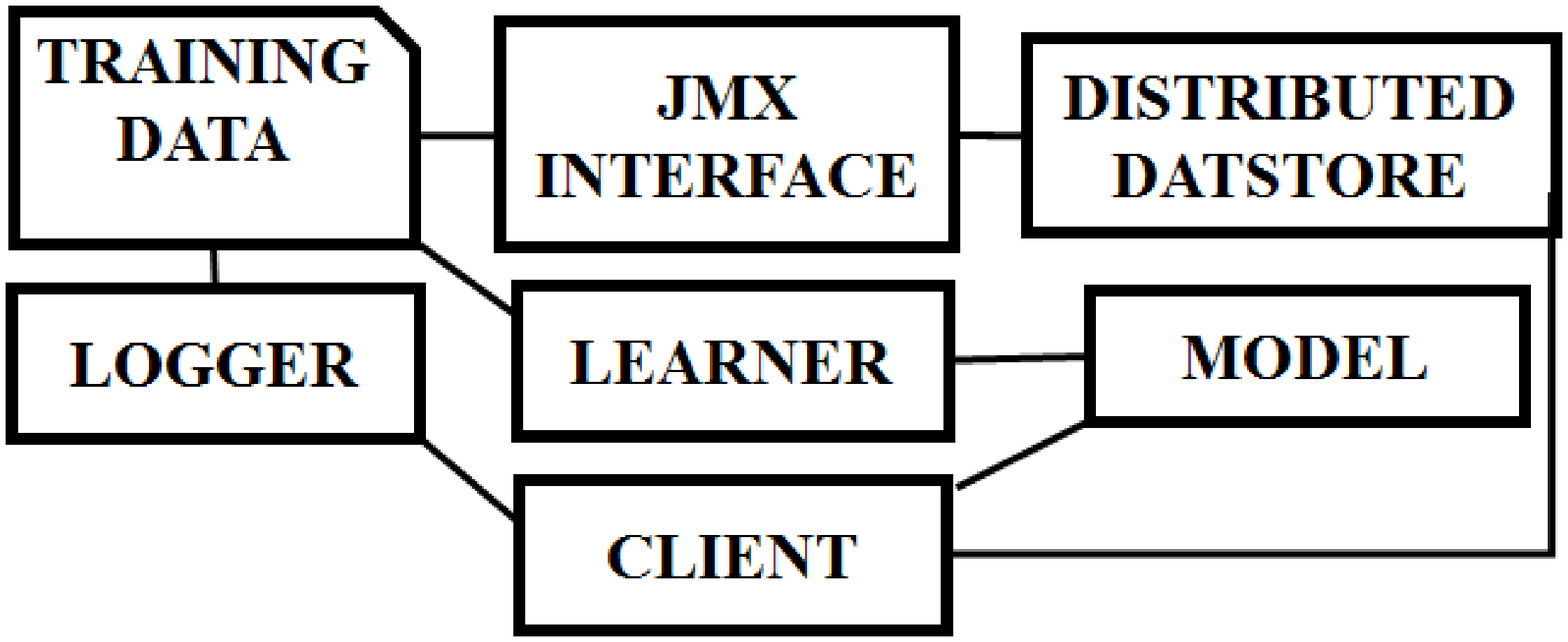}
        \caption{The architecture of OptCon: rectangles denote modules and the folded rectangle denotes dataset.}
        \label{fig:Arch}
\end{figure}
 %\begin{figure}[!htb]
%\centering
%%\noindent
%\subfigure[The Architecture of OptCon]{\includegraphics[width=0.55\linewidth,height=0.7in]{arch}\label{fig:Arch}}\hfill
%\subfigure[Example subSLAs]{\scalebox{0.85}{\begin{tabular}[b]{|c c|} % centered columns (4 columns)
%\hline\hline %inserts double horizontal lines
% Average Latency & Staleness \\ % inserts table heading
% \hline\hline
% $\leq$100ms & $\leq$5ms \\ % inserting body of the table
% $\leq$50ms & $\leq$10ms \\
% $\leq$25ms & $\leq$15ms \\ % [1ex] adds vertical space
%\hline %inserts single line
%  %\includegraphics[width=1.5in,height=0.7in]{arch}
%\end{tabular}\label{table:sla}}}
%\caption{Architecture and subSLA of OptCon}
%\end{figure}
%Figure \ref{fig:Arch} describes the architecture of OptCon.
 OptCon (refer to Figure \ref{fig:Arch}) consists of the following modules:
%\begin{enumerate}
    1) The Logger module records the independent variables (Section \ref{sec:autocase}), the observed latency, and staleness, obtained using experiments performed on the datastore with benchmark workloads. It collates these parameters into a training data corpus, which acts as the training dataset. % for training the model, and  %using the Java Management Extensions (JMX) Interface \cite{Lakshman:2010:CDS:1773912.1773922} provided by the storage system,
    2) The Learner module learns the model $\mathcal{M}$ (refer Section \ref{sec:over}) from the training dataset, and predicts a matching consistency level that maximizes the throughput.
    3) The Client module calls the other modules, and executes the given operation on the datastore.
    %3) a Client module that predicts the matching consistency level, and performs the requested operation by calling the query client.
%\end{enumerate}
%\marginpar{briefly describe the functionalities of the three components 1 sentence for each.}
  %The JMX Interface  %is an extension over YCSB 0.1.4 \cite{Cooper:2010:BCS:1807128.1807152}, running
  \par During the training phase, the Client module runs a simulation workload on the given quorum-based datastore. The Client calls the Logger module (Figure \ref{fig:Arch}) to collect the independent variables (parameters) and the observed parameters %, i.e., latency, staleness, and throughput,
   for the operation from the JMX interface \cite{Lakshman:2010:CDS:1773912.1773922}, %, built-in with the distributed datastore.
   and appends these parameters into the training data. The Client then calls the Learner module, which trains the model $\mathcal{M}$ from the training data applying machine learning techniques.
   %Subsequent operations on the distributed datastore are processed through the Client.
    During the prediction phase, the Client calls the Learner, with runtime arguments comprising the subSLA thresholds and the current values of the independent variables.
   The subSLA (and also the SLA) can be varied on a per-operation basis. The Learner predicts a matching consistency level from the learnt model $\mathcal{M}$. The Client performs the given operation on the datastore with the predicted consistency level.

%Subsubsection text here.

\subsection{Design of the Logger Module}\label{sec:logdes}
 %The Logger module is designed as an extension over the YCSB 0.1.4 \cite{Cooper:2010:BCS:1807128.1807152} framework
 % The Logger module (Figure \ref{fig:Arch}) computes the training parameters for learning the model $\mathcal{M}$, and collates these parameters into a training data corpus.
%  The Logger module runs as a daemon on top of the Distributed Datastore, listening for the parameters using the JMX Interface.
  %It also performs the required pre-processing on the collected parameters, and computes the parameters, like client-centric staleness. %In Section \ref{sec:logparam}, we discuss the parameters collected by the Logger. Section \ref{sec:logstale} discusses the details of the client-centric staleness metric used in OptCon. %The parameters are collated into Training Data, i.e., the training dataset file.

  \subsubsection{Model Parameters Collected by the Logger}\label{sec:logparam}
  %Here we describe the parameters that the Logger collects for building the training data.
   For the reasons explained in Section \ref{sec:parameters}, the latency $L$, i.e., time delay for the operation, %the throughput $T$, %(refer Table \ref{table:glossary}),
 and client-centric staleness $S$, %i.e., the observed consistency,
%%throughput,
% \begin{itemize}
% \item operation latency $L$ and
% \item average staleness $S$
%  \end{itemize}
    are considered as the model parameters that need to satisfy a given SLA. The throughput $T$ is considered as a secondary optimization criterion. As discussed in Section \ref{sec:parameters}, server centric parameters, like replication factor and keyspace size, are not considered. Following the reasoning in Section \ref{sec:autocase}, the independent variables (parameters) for learning the model $\mathcal{M}$ are: 1) the read proportion ($\mathit{RW}$) in the operation, 2) the thread count ($\mathit{Tc}$), i.e., the number of user threads spawned by the client, 3) the packet count ($P$), i.e., the number of network packets  transmitted during the operation, and 4) the consistency level $C$, provided as a categorical attribute containing attribute values specific to the given datastore. %A given value of $C$ is given as the nominal. %represents a specific number of replicas that require to respond under the consistency level given by $C$. % (refer Table \ref{table:glossary}).
     \par Most operations on quorum based stores are not instantaneous, but are executed over certain time intervals \cite{DBLP:conf/cloud/GolabRAKWG13}. Following  \cite{Bailis:2012:PBS:2212351.2212359}, the Logger uses the average latency over a time  interval of one minute for measuring $L$.
%    \begin{itemize}
%    \item thread count, i.e., number of the client threads,
%    \item the packet count, and
%    \item read-write proportions.
%     \end{itemize}
      %Client-centric staleness $S$ is measured as the timestamp difference between the current operation on an item and the latest update to this item.   %It depends on the replication factor and
%topology.
  The Logger computes $S$ in terms of the {\boldmath$\Gamma$} metric of Golab et al. \cite{DBLP:conf/cloud/GolabRAKWG13}. As demonstrated in the above work, {\boldmath$\Gamma$}  is preferred over other client-centric staleness measures for its proven sensitivity to workload parameters and consistency level. %, as demonstrated in \cite{DBLP:conf/cloud/GolabRAKWG13}. %, owing to the reasons given later in Section \ref{sec:logstale}
 Section \ref{sec:sig} establishes the significance of the above parameters with respect to the OptCon model.
  Section \ref{sec:compare} demonstrates the accuracy of the model comprising the that the above parameters. % produce high prediction accuracy.
%The Learner module learns a model from the Training Data, consisting of the above variables collected by the JMX Interface and collated by the Logger module, and predicts matching client side consistency settings under a given subSLA.

\subsubsection{Computation of $\Gamma$ as the Metric for Client-centric Staleness}\label{sec:logstale}
   The metric $\Gamma$ %\cite{DBLP:conf/cloud/GolabRAKWG13}
    is based upon Lamport's \emph{atomicity} property \cite{lamport_atomic}, which states that operations appear to
	take effect in some total order that reflects their ``happens before'' relation in the following sense:
	if operation A finishes before operation B starts then the effect of A must be visible before the effect of B.
We say that a trace of operations recorded by the logger is {\boldmath$\Gamma$}-atomic if it is atomic in Lamport's sense, or becomes atomic
after each operation in the trace is transformed by decreasing its starting time by {\boldmath$\Gamma$}/2 time units, and increasing
its finish time by {\boldmath$\Gamma$}/2 time units.
 In this context the start and finish times are shifted in a mathematical sense for the purpose of analyzing a trace, and do not imply injection of artificial delays;
the behavior of the storage system is unaffected.
%The $\Gamma$ metric can be defined at various granularity levels, including
    \par The \emph{per-key $\Gamma$ score} quantifies the degree of client-centric staleness incurred by operations applied to a particular key. %We use the $95^{th}$ percentile per-key $\Gamma$ score as the client-centric staleness measure. We order the per-key $\Gamma$ scores observed in a trace from lowest to highest and select the $(0.95*n)^{th}$ per-key $\Gamma$  value as the $95^{th}$ percentile per-key $\Gamma$, where $n$ is the total number of keys accessed in the trace.
     We considered average per-key  $\Gamma$, but settled for percentile per-key  $\Gamma$ since it takes into account the skewed nature of the workloads.  %The averaging is used to smooth out the noise in the results, although it may not work best on skewed workload. We adopt as our measure of staleness an \emph{average {\boldmath$\Gamma$} score} defined as follows: $$ \frac{\sum \mbox{per-key } \Gamma \mbox{ scores}}{\mbox{total \# of keys accessed in the trace}}$$

\subsection{Design of the Learner Module}\label{sec:approach}
 %In the absence of a closed form mathematical model \cite{Bailis:2012:PBS:2212351.2212359} formally relating the parameters---client-side consistency, observed
% latency, and staleness---we use machine learning  to learn a model from historic data.
%We discuss the design of the Learner module (refer to Figure \ref{fig:Arch}), which learns a
%%decision tree \cite{Quinlan:1986:IDT:637962.637969} based
%model from the training data.
 %The Learner module (refer to Figure \ref{fig:Arch}) classifies the training data according to a matching consistency levels with respect to the values of the model parameters in each row. %We evaluate various learning algorithms for OptCon using a group of model selection metrics --- the developers can use our analysis to decide on the appropriate learning technique to use for a given distributed datastore.
   %The Client module predicts the consistency level that produces the highest throughput under a given set of thresholds specified/ in the subSLAs.  %Feasibility here refers to the constraint that the consistency level so chosen
%should not result in bringing performance metrics to levels so high (or so low) such that it disrupts the
%normal flow of traffic beyond a threshold.
%If there are multiple candidate consistency levels that give values of dependent variables satisfying the
%thresholds, we choose the strongest consistency level. This
% ensures that the user gets the best possible results for the particular use case in a given network state,
% without violating the thresholds specified in the subSLA.
In OptCon, building the model $\mathcal{M}$ from training examples (Figure \ref{fig:Arch}) is a one-time process, i.e., the same model can be reused for predicting consistency levels for all operations. Even with node failures, the model may not need to be regenerated since the training data remain unchanged, only the failed predictions, resulting due to node failure, need to be rerun.
 However, the model will need to be retrained in certain cases where the behavior of the system with respect to
 the latency and staleness changes, such as when a storage system receives a software upgrade. %Next Section \ref{sec:strategies} discusses the various candidate learning strategies for implementing the Learner module.
\subsubsection{Overview of Possible Learning Techniques}\label{sec:strategies}
%\begin{table}[!htb]
%   % \begin{table}
%  % title of Table
%%\centering % used for centering table
%\scalebox{0.9}{
%\begin{tabular}{|c|c|c|c|c|} % centered columns (4 columns)
%\hline\hline %inserts double horizontal lines
%\bf Approach & \bf Cross Validation Error & \bf AICC & \bf BIC & \bf Overhead (ms) \\ % inserts table heading
% \hline\hline
%Decision Tree & $0.14$ & $10.73$ & 51.44 & 1 \\ \hline
%%SVM & $0.44$ & $11.21$ & 51.93 & Low \\ \hline % [1ex] adds vertical space
%Bayesian Learning & $0.57$ & $12.85$ & 53.57 & 1.2 \\ \hline % [1ex] adds vertical space
%Logistic Regression & $1.98$ & $16.32$ & 57.07 & 0.7 \\ \hline % inserting body of the table
%Random Forest & $0.14$ & $9.51$ & 50.24 & 1.3 \\  \hline% [1ex] adds vertical space
%Neural Network & $0.059$ & $12.85$ & 63.54 & 1.5 \\ % [1ex] adds vertical space
%\hline %inserts single line
%\end{tabular}
%}
%% is used to refer this table in the text
% \caption{Model Selection Results}
%%\end{table}
%\label{table:selection}
%\end{table}
 %Here we consider various possible machine learning algorithms that can be used to implement the Learning module.
  We provide a brief overview of the possible Learning techniques and their applicability to OptCon. We describe the implementation details, including the various configuration parameters, of each learning algorithm in Section \ref{sec:impl}. Performance analysis and insights gathered from each technique are given in Section \ref{sec:compareresults}.  %OptCon acts as an interface between client application and the distributed datastore. Hence, speed and simplicity are important criteria, apart from accuracy, for the choice of learning techniques in OptCon as we wish to minimize the latency overhead for learning the model.
   %We list down the measured speed and accuracy for each algorithm in the evaluation section (refer section \ref{sec:eval}).
    We consider Logistic regression %\cite{Andrew:2007:STL:1273496.1273501}
 and Bayesian learning as they can help  visualize the significance of the model parameters and the dependency relations among these parameters, and thus can provide intuition to the developer for developing complex and more accurate models \cite{Flach:2012:MLA:2490546}.
  %Linear regression failed to produce an acceptable model - because of the absence of linear dependence among the variables.
  We also consider Decision Tree, Random Forest, and Artificial Neural Networks (ANN), %\cite{Flach:2012:MLA:2490546},
   since they can produce more accurate predictions, being directly computed from the data.
    We consider these approaches in the order of their performance given in Table \ref{table:selection}, in terms of various model selection metrics. %\cite{Flach:2012:MLA:2490546}.
 %The approach is to favor light weight learning algorithms over compute intensive techniques, while not compromising on accuracy, since compute intensive algorithms generally sacrifices response time to provide higher accuracy.
  We leave the choice of a suitable learning technique to the developer, rendering flexibility to the framework. Based on our evaluation of the learning algorithms, developers can choose the learning technique that best suits the respective application domain and use case.  %, making it adaptable to other Dynamo-style systems.
%In the evaluation section (refer section \ref{sec:eval}), we compare the performance of each of the techniques on Cassandra to enable the developers to make an informed choice.
\par %Here we discuss each learning algorithm and the possibility of its application to OptCon.
%We initially started our analysis with Linear regression but rejected it, since it produced near random prediction accuracy. The failure of linear regression can be attributed to: 1) too much simplicity in  assuming linear model, and 2) our problem is more of a classification problem as explained in the next paragraph.   %We start with the logistic regression \cite{Andrew:2007:STL:1273496.1273501} technique.
%We divide the latency and staleness values into ranges and denote each range as a category. We express latency $L$  and staleness $S$ as the dependent variables of two objective functions of logistic nature, comprising independent variables $RW$, $Tc$, and $P$.
  With logistic regression %\cite{Flach:2012:MLA:2490546} %\cite{Andrew:2007:STL:1273496.1273501}
   approach, we fit two logit (i.e., logistic) functions for the two dependent variables  $L$ and $S$ as follows: $\mathit{logit}\left( \pi \left( L \right) \right) = \beta_0 + \beta_1 C + \beta_2 RW + \beta_3 P + \beta_4 Tc$ and $\mathit{logit}\left( \pi \left( S \right) \right) = \beta_4 + \beta_5 C + \beta_6 RW + \beta_7 P + \beta_8 Tc$, where $C$ is the applied consistency level, $L$ is the observed latency, $S$ is the observed staleness, $RW$ is the proportion of reads, $P$ is number of packets transmitted, and $Tc$ is the number of threads during an operation. % , where the consistency level $C$ is given as a categorical variable.
    $\beta_i$ are the coefficients estimated by iterative least square estimation, and $\pi$ is the likelihood function. Using ordinary least square estimation, we iteratively minimize a loss function given by the difference of the observed and estimated values, with respect to the training dataset. For eliminating overfitting, we perform $L^1$ regularization with the Lasso method.  %\cite{Flach:2012:MLA:2490546}.
 %Linear Regression is a particular form of regression where the dependent variable $y$ varies linearly with
%respect to the independent variable $x$.
%Linear Regression can be used to make prediction using simple linear models fitted on a dataset of observed values for given parameters. Assuming linear dependency, we start with the linear regression technique in an attempt to come up with a simple mathematical model to express the effect of the various controlling parameters $RW$, $Tc$, and $P$, on average latency and staleness for an operation, under different client side consistency levels $C$. Also, linear regression can be used to determine the strength of the relationship of the individual independent variables on the dependent variables, and act as a basis for further complicated modelling algorithms. Thus, we use ANOVA analysis on linear regression to determine which controlling parameters from $RW$, $Tc$, and $P$ truly determine the resultant variations in staleness and average latency, under the different client side consistency settings.
 Next we consider decision tree learning algorithm because of its simplicity, speed, and accuracy.
The given problem can be viewed as a classification problem, which classifies the training dataset into
classes based on whether the observed latency and staleness (corresponding to each row) fall within the given
SLA thresholds. In fact, the problem statement, i.e., choosing a matching consistency level, can be viewed as
a classification problem. The modelling technique using Decision Tree comprises the following phases: 1) \textbf{Labelling}: %Normally,
%the secondary optimization criteria (i.e., throughput $T$)  would be dealt with only after the primary SLA parameters
%are satisfied, i.e., during the prediction phase. However,
On the fly computations for maximizing $T$ (i.e., the secondary optimization criteria) in the prediction phase would introduce
considerable overhead. Since the latency is considered before $T$ is maximized, the additional overhead may result in violation of the latency threshold in the SLA. Hence, OptCon performs the computations for $T$ in the labelling phase itself.  Using exhaustive search, we label each row in the training dataset with the highest $T$ corresponding to a given values of $\mathit{RW}$, $\mathit{Tc}$ and $P$, such that the SLA parameters $L$ and $S$ are within the given SLA thresholds. 2) \textbf{Training}: Next we apply decision tree learning for training a model $\mathcal{M}$ from the labelled dataset. Use of error pruning techniques mitigates issues of overfitting \cite{Flach:2012:MLA:2490546}.
 Next, we consider the random forest %\cite{Flach:2012:MLA:2490546}
 algorithm that applies ensemble learning to bootstrap multiple weak decision tree learners to generate a strong classifier. It iteratively chooses random samples with replacement from the training dataset, and trains a decision tree on each of the samples. We take the average of the predictions from each decision tree to obtain the final prediction with respect to a test dataset.
  For the Bayesian %\cite{Flach:2012:MLA:2490546}
   approach, we make the following simplifying assumptions to make our problem suitable for Bayesian learning: A1) a prior logistic distribution exists for each of $RW$, $Tc$, $P$, and $C$ (see Section \ref{sec:design}), A2) the
 posterior distributions for
    $L$, and $S$ are conditionally dependent on the joint distribution of $\tuple{RW,\; Tc,\; P,\; C}$, and A3) the target features $L$ and $S$ are conditionally independent.
     With these approximations, we plug-in the resultant dependency graph from the Logistic Regression approach as input to the Bayesian learner. Despite these assumptions and approximations, the Bayesian technique can provide intuition to the developers, that can be used to build a more accurate model. % a mathematical basis for applying complex algorithms.
      %Next, we try Artificial neural networks \cite{Flach:2012:MLA:2490546} approximates an activation function, a weighted nonlinear function,  by updating the weights through iterative backpropagation-based learning process. SVM, i.e., Support Vector Machine \cite{Flach:2012:MLA:2490546}, performs training on the training dataset to define a separation boundary, in form of a hyperplane, to classify the dataset into.
      Next, we consider Artificial Neural Networks (ANN) %\cite{Flach:2012:MLA:2490546} %and Support Vector machine (SVM)
      for learning a model $\mathcal{M}$. %ANN can provide high accuracy with multiple iterations. %, thus causing high training overhead.
       Though the training phase for ANN is not a factor (since training occurs once), the overhead for the prediction phase is still considerably high (Section \ref{sec:over}) due to model complexity.
       %We do not discuss SVM, since it requires an appropriate kernel function. In the absence of a closed-form mathematical model, we are unable to define an exact kernel function. %Hence we get unacceptably high error (hence we omitted the results).
%The conclusion goes here.

\section{Implementation and Evaluation}\label{sec:eval}
%We compare the results obtained with various modelling techniques, for the sake of exploring and evaluating the different methods and providing a guideline to future researchers and developers.
 %We use YCSB \cite{Cooper:2010:BCS:1807128.1807152} as a client for performing operations on Cassandra. %Cassandra is an open source key value based distributed data store. It
%is widely used in the industry as well as academic world for its scalability, partition tolerance, and
%availability. %However, OptCon has been built as a Java-based wrapper, following a loosely-coupled
%architecture and may have the potential to be used with other distributed storage systems with some modifications.
     %We present the implementation and evaluation details of OptCon.
      %In Section \ref{sec:setup}, we describe the experimental setup. Section \ref{sec:experimental} discusses the threshold ranges used in the subSLAs. Section \ref{sec:over} quantifies the OptCon overhead. Section \ref{sec:impl} discusses implementation details of the OptCon modules, %i.e., the Logger module and the Learner module,
%      including the configuration parameters used for the different learning algorithms. Section \ref{sec:sig} performs dimensionality reduction and gives the significance of each independent variables in the model. Section \ref{sec:compare} presents comparative study of the performance of the various learning algorithms used. Section \ref{sec:vary} demonstrates the adaptability of OptCon to changing workload. Section \ref{sec:user} demonstrates the adaptability of OptCon to different subSLAs.

\subsection{Experimental Setup}\label{sec:setup}
Following \cite{DBLP:conf/cloud/GolabRAKWG13}, we have run our experiments on a testbed of 20 Amazon Ec2 small
instances, located in the same Ec2 region, running Ubuntu 13.10, loaded with Cassandra %\cite{Lakshman:2010:CDS:1773912.1773922}
 with a replication factor of 5.  Our models are learnt from a training dataset comprising 23040 rows, and a testing dataset of 2560 rows, generated by running varying YCSB \cite{Cooper:2010:BCS:1807128.1807152} workloads.
%Cassandra nodes connected over a private intranet.
%We have developed the Logging module as an extension of the widely recognised YCSB \cite{Cooper:2010:BCS:1807128.1807152} benchmark tool
%(version 0.1.4) running on top of Cassandra 2.1.0.
 %The cluster consists of multiple disjoint nodes connected by a
%network with lags/time delays in the individual system clocks.
% Without time synchronization among the various nodes, the calculated measures like latency would be rendered
%useless because of the clock skew.
 We have used the NTP (Network Time Protocol) protocol %\cite{Mills:1992:NTP:RFC1305}
 implementation from ntp.org to bring down the maximum clock skew to 5 ms. %This
%accounts for a margin of 5 ms for error in the measurement. %, which is further reduced with averaging on multiple observations.

\subsection{Latency and Staleness Ranges Used in the subSLAs}\label{sec:experimental}
 Studies \cite{Everts:2012:Online}
         suggest that
 the latency for standard web applications falls in the range of 75-140 ms. The Pileus system
\cite{Terry:2013:CSL:2517349.2522731}, which includes latency in subSLAs as well, uses values between  200 to 1000 ms as the latency
threshold. Further reducing the lower bound to favor availability, we choose any value between 101 and 150ms as a candidate for the threshold of latency in OptCon. Following the observations from the Riak deployment at Yammer \cite{YMMRRiak}, Bailis et al. \cite{Bailis:2012:PBS:2212351.2212359} uses a range of 1.85-13.6 ms for the staleness bounds, given as the
        t-visibility values. Hence, following \cite{Bailis:2012:PBS:2212351.2212359}, we choose a value within the above range as the staleness threshold. Rounding the bounding values for the ranges given above, our subSLAs use values in the ranges  101-150 ms and 1-13 ms as thresholds for $L$ and $S$, respectively.

\subsection{OptCon Framework Overhead}\label{sec:over}
Training the model being a one-time process, the latency overhead due to the training phase is negligible during an operation. However, there is still considerable overhead due to the prediction phase, especially with a large training dataset. The  prediction phase spans from the call to the Learner module till the predicted consistency level is returned by the Learner module. The average, variance and standard deviation of the overhead are 1 ms%2.12 ms
        , 0.25, and 0.37, respectively, with decision tree learning. The average overhead with
         different learning techniques is given in Table \ref{table:selection}.

\subsection{Implementation of the Modules: Logger and Learner}\label{sec:impl}
OptCon is developed as a Java-based wrapper over Cassandra v2.1.0.  %\cite{Lakshman:2010:CDS:1773912.1773922}.
The source code can be found in the github
repositories: \url{https://github.com/ssidhanta/YCSBpatchpredictconsistency/}, \url{https://github.com/ssidhanta/TrainingModelGenerator/}, and \url{https://github.com/ssidhanta/HectorCient/}.
The Logger module is implemented as an extension over the YCSB 0.1.4 \cite{Cooper:2010:BCS:1807128.1807152} framework. It runs as a daemon on top of the datastore, listening for the parameters using the built-in JMX Interface.
%\subsection{Implementation Details of the Learning Module Using Various Learning Techniques}\label{sec:impl}
   %Next we describe the implementation details of the Learner module. %, including the tools used, and the configuration parameters.
   Logistic Regression, SVM, and Neural Network implementations of the Learning module use Matlab 2013b's statistical toolbox. %\cite{MATLAB:2013}
 %to implement the Learner module.
 Decision Tree and Random Forest are implemented using Java APIs from Weka, an open source machine learning suite.
 %\cite{Hall:2009:WDM:1656274.1656278}.
  %Weka is an open source machine learning suite implementing a host
%    of machine learning algorithms.
 The Bayesian approach uses Infer.net, %\cite{InferNET14},
  an open source machine learning toolbox under the .NET framework.
   We summarize the important configuration parameters in the Learner implementations. Logistic regression uses a minimization function with additional regularization weights, multiplied by a tunable $L^1$ %(i.e., Lasso)
  norm term $\Lambda$. %\cite{Flach:2012:MLA:2490546}.
   We set  $\alpha$ = 0.05 and $\Lambda$ = 25.
  Decision tree uses a confidence threshold  of 0.25 for pruning, and uses random data shuffling with seed = 1. %\cite{Flach:2012:MLA:2490546}.
   Random forest  bootstraps 100 decision trees, and introduces low correlation  among the individual trees by selecting a random subset of the features at each split boundary of the component trees.  %\cite{Flach:2012:MLA:2490546}.
    The Bayesian method uses a precision of 0.1. %\cite{Flach:2012:MLA:2490546}.
     The ANN is implemented as a two-layer perceptron, with default weights = 0 and the  default bias = 0. %For SVM, in the absence of a concrete mathematical basis for the model, we tried out a linear kernel for its simplicity.

\subsection{Dimensionality Reduction: Significance of the Model Parameters}\label{sec:sig}
\begin{table}[!htb]
   % \begin{table}
  % title of Table
%\centering % used for centering table
\scalebox{0.9}{
\begin{tabular}{|c|c|c|c|}
\hline\hline
{\bf \textbf Parameter} & {\bf \begin{tabular}[c]{@{}c@{}}\textbf Rank Features \\ Technique\end{tabular}} & {\bf \begin{tabular}[c]{@{}c@{}}\textbf Sequential Attribute \\ \textbf Selection Technique\end{tabular}} & {\bf \begin{tabular}[c]{@{}c@{}}\textbf Misclassification \\ Error For Random Forest\end{tabular}} \\ \hline \hline
C             & 4.81                                                                     & 1                                                                                       & 0.22                                                                                       \\ \hline
RW            & 0.9                                                                      & 1                                                                                       & 1.11                                                                                       \\ \hline
P             & 2.78                                                                     & 1                                                                                       & 0.24                                                                                       \\ \hline
\end{tabular}
}
% is used to refer this table in the text
 \caption{Significance of the Model Parameters: As per descending order of Rank Features and Sequential Attribute Selection, and ascending order of Misclassification error}
%\end{table}
\label{table:feature}
\end{table}

%We evaluate the significance of each  model parameter with respect to the model.
 We apply dimensionality reduction \cite{Flach:2012:MLA:2490546} to determine the features (parameters) that are relevant to the problem, and have  low correlation among themselves. We use the following techniques to detect any redundant feature that can be omitted without losing information. %, applying: 1) rank features, 2) sequential feature selection, or 3) misclassification error techniques. %, using either filtering or wrapper approaches. %We include the features with low correlation using either filtering or wrapper approaches.
%We rank features applying filtering technique as a preprocessing step before actually performing training.
   Rank features %\cite{Flach:2012:MLA:2490546}
    obtains the most significant features by computing  different criterion functions. It is applied as a filtering (or a preprocessing) step before the training phase. %We use functions like the t-test parameter, Kullback-Leibler distance, AUC measure for the ROC curve, Bhattacharya parameter, and u-statistic obtained from the Wilcoxon test.
 On the other hand, sequential feature selection %\cite{Flach:2012:MLA:2490546}
  is a wrapper technique, that sequentially selects the features until there is no substantial improvement (given by the deviance of the fitted models) in the predictive power for an included feature. For the random forest model, we evaluate the significance of the features in the ensemble of component trees using the misclassification error. %\cite{Flach:2012:MLA:2490546}.
 %Then, the features are ordered in an increasing order of significance based on the magnitude of the misclassification error.
 Table \ref{table:feature} presents the results of the above techniques on the model parameters  $C$, $RW$, and $P$. %According to the results for sequential feature selection (refer to Table \ref{table:feature}), $C$, $RW$, and $P$ are all significant for the model
 Sequential feature selection outputs 1 for all the model parameters, indicating  that all the parameters are significant. Rank features ranks the relative significance of the features in the order $C$, $P$, and $RW$. Misclassification error ranks the features in the order $RW$, $P$, and $C$. The order of the relative significance is different with each dimensionality reduction technique; the developer must choose the suitable ranking approach based on the learning technique. For example, the misclassification error criterion is to be used for the random forest technique.

\subsection{Comparison of the Predictive Power of the Learning Techniques Using Model Selection Metrics}\label{sec:compare}

\begin{figure*}[htpb]
\centering
%\noindent
\subfigure[Operations With Consistency Levels Predicted by OptCon satisfy the subSLA in $100$\% cases]{ \includegraphics[width=0.3\linewidth,height=4.2cm]{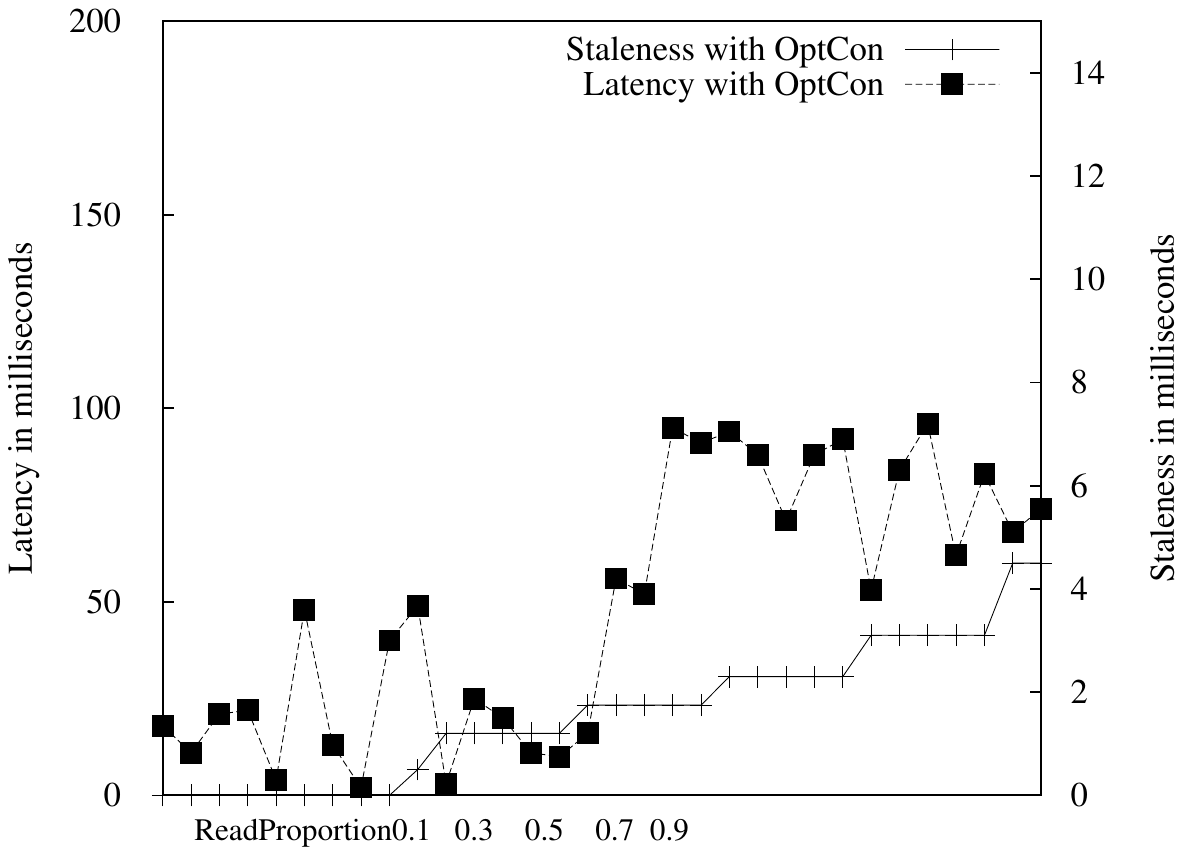}\label{fig:label-15}}\hfill
\subfigure[Operations With READ ALL/WRITE ALL satisfy the subSLA in $70$\% cases]{\includegraphics[width=0.3\linewidth,height=4.2cm]{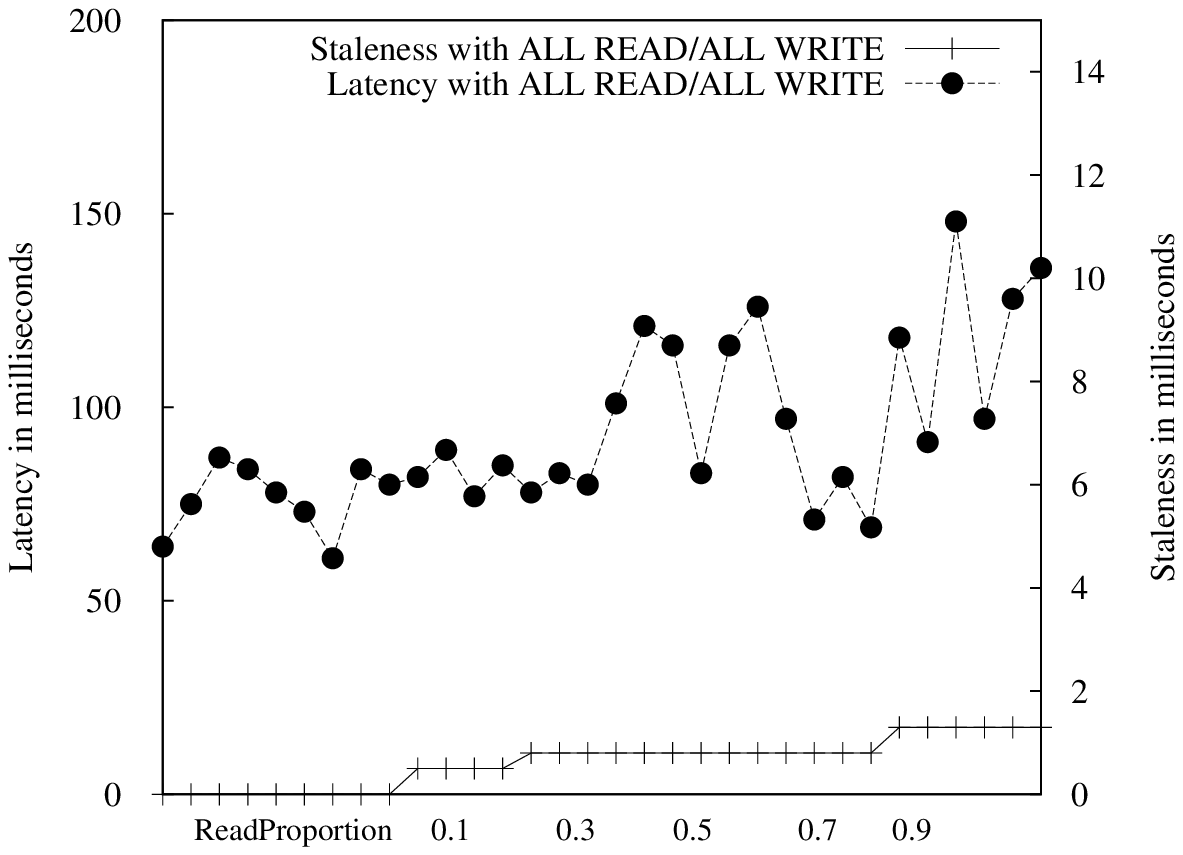}\label{fig:label-4}}\hfill
\subfigure[Operations With READ ALL/WRITE QUORUM satisfy the subSLA in $75$\% cases]{\includegraphics[width=0.3\linewidth,height=4.2cm]{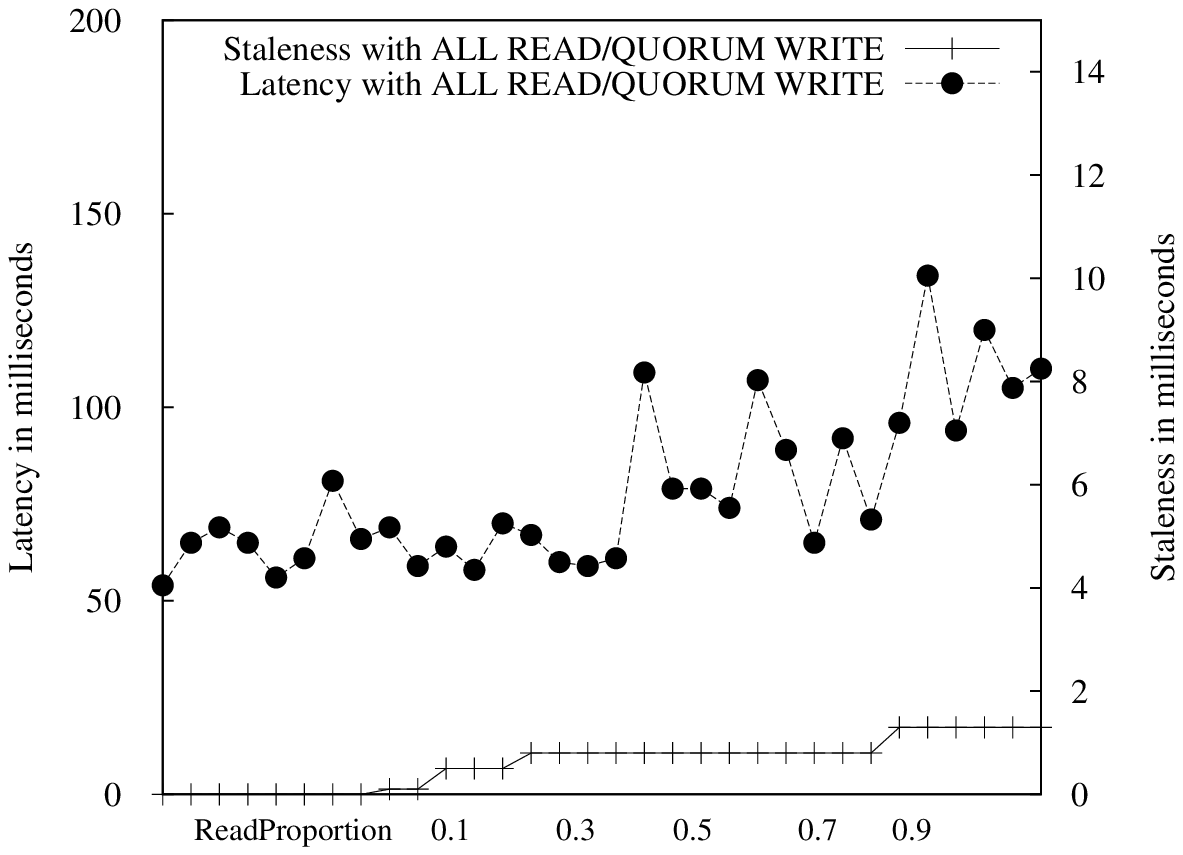}\label{fig:label-5}}\hfill
\subfigure[Operations With READ QUORUM/WRITE QUORUM satisfy the subSLA in $75$\% cases]{ \includegraphics[width=0.3\linewidth,height=4.2cm]{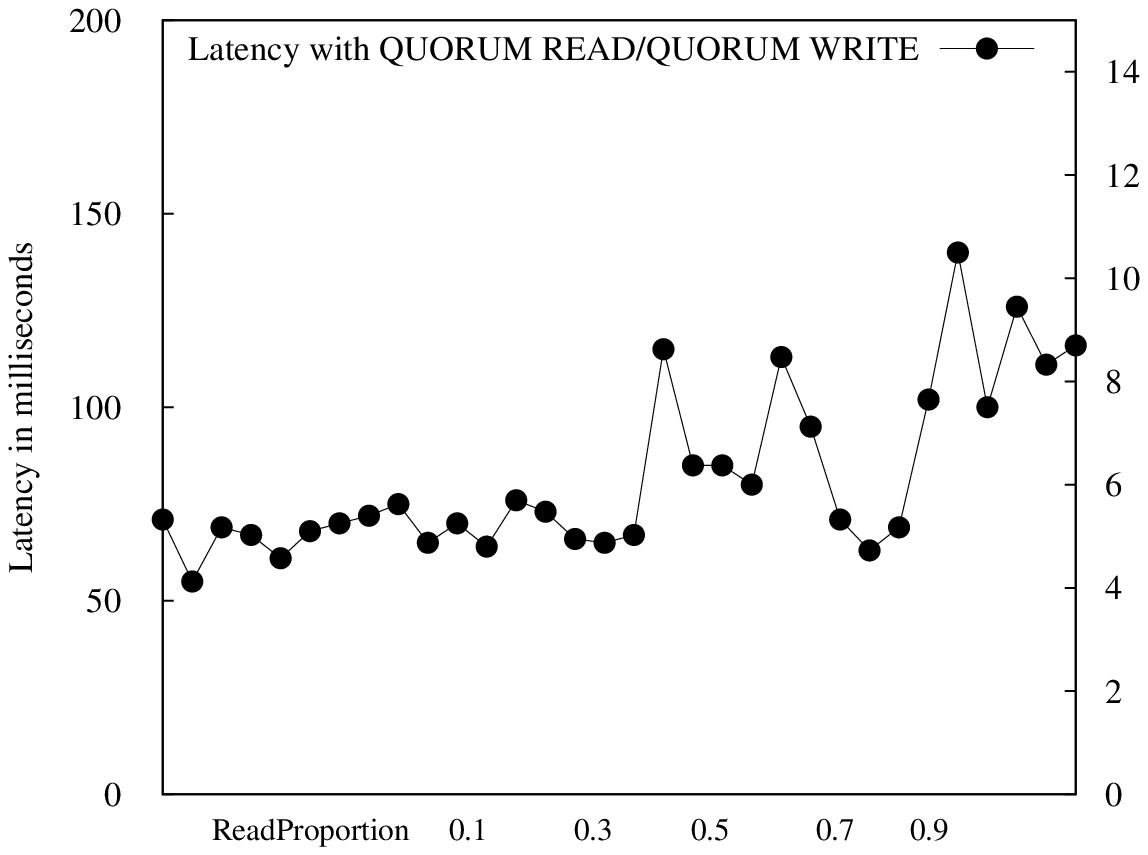}\label{fig:label-6}}\hfill
\subfigure[Operations With READ QUORUM/WRITE ALL satisfy the subSLA in $35$\% cases]{ \includegraphics[width=0.3\linewidth,height=4.2cm]{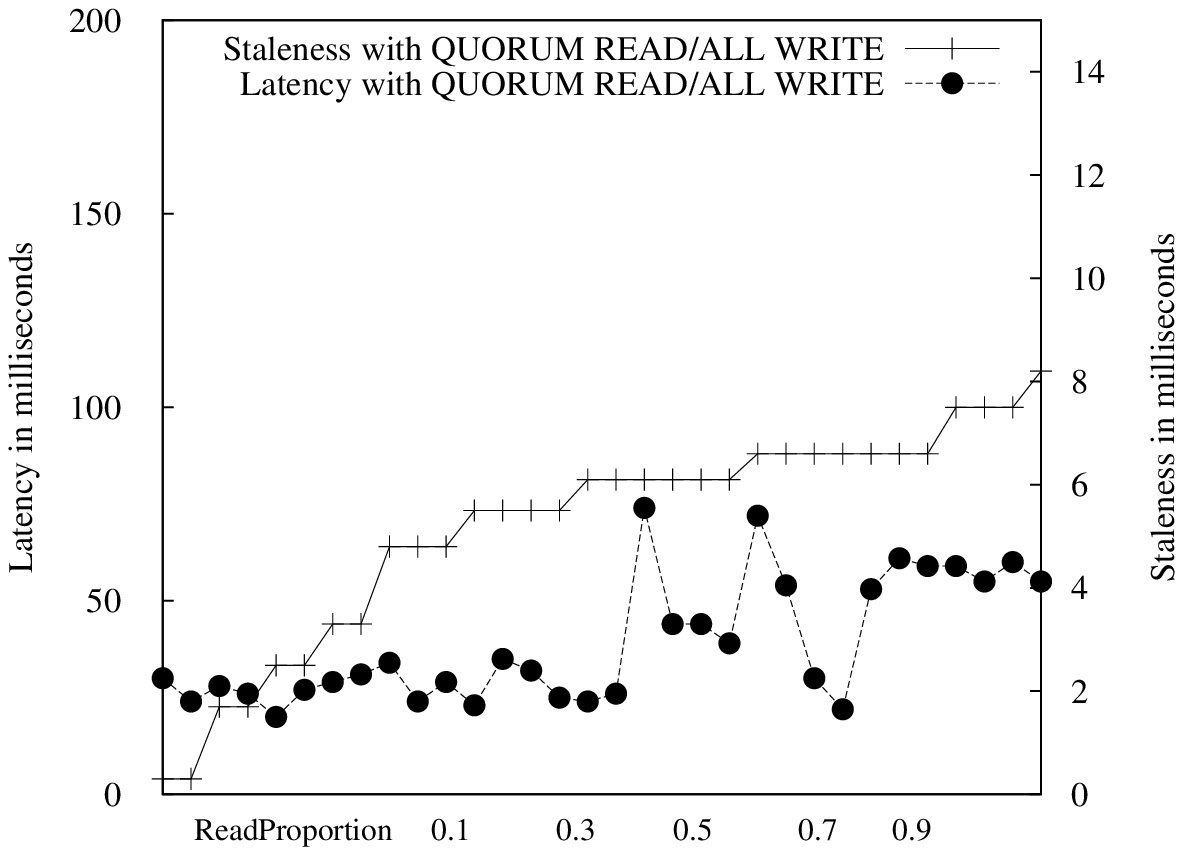}\label{fig:label-7}}\hfill
\subfigure[Operations With READ ALL/WRITE ANY satisfy the subSLA in $30$\% cases]{ \includegraphics[width=0.3\linewidth,height=4.2cm]{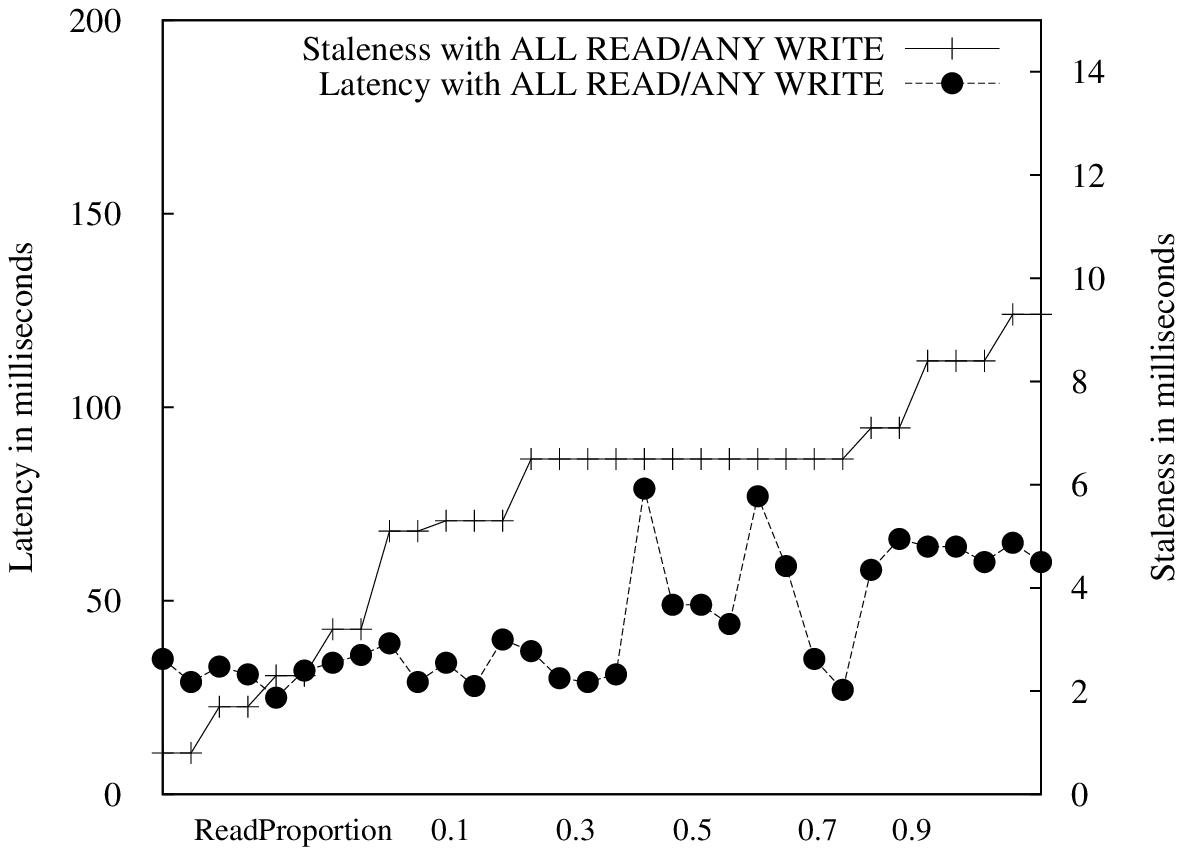}\label{fig:label-8}}\hfill
\subfigure[Operations With READ ALL/WRITE ONE satisfy the subSLA in $30$\% cases]{ \includegraphics[width=0.3\linewidth,height=4.2cm]{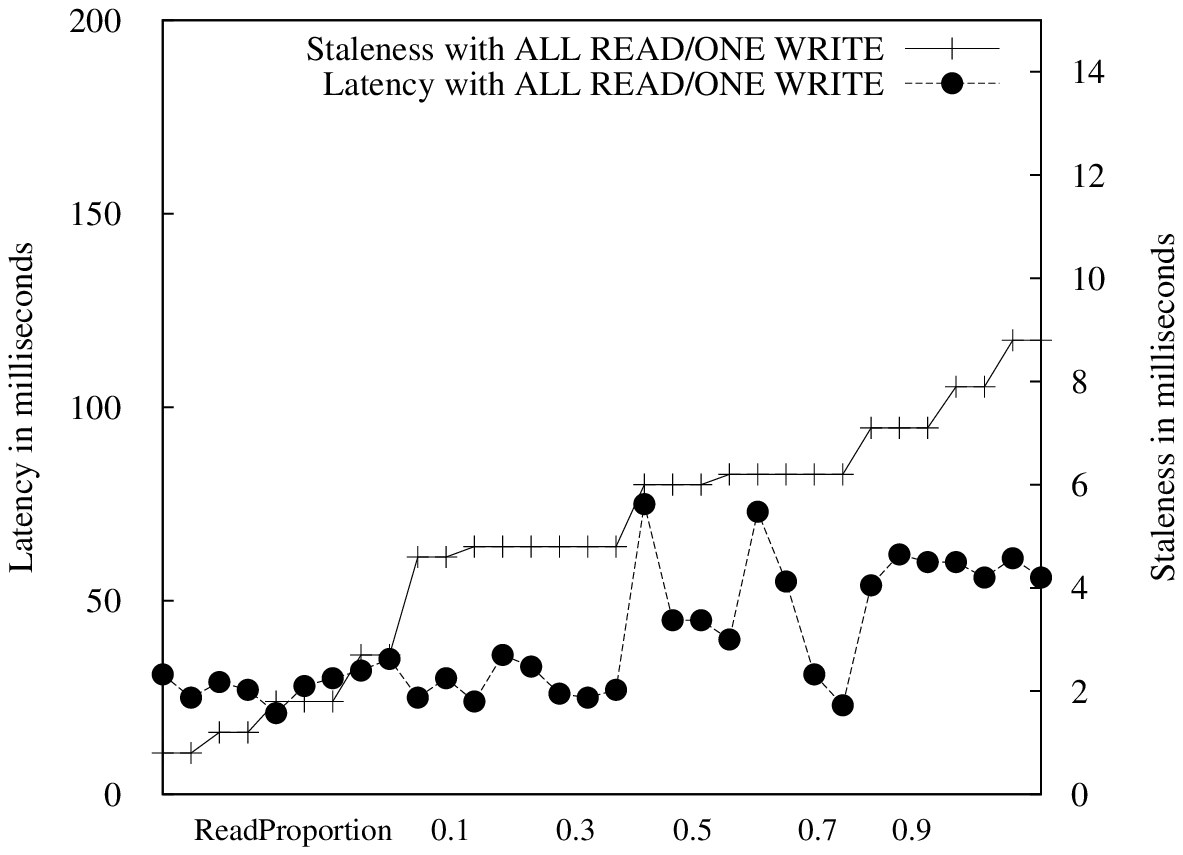}\label{fig:label-9}}\hfill
\subfigure[Operations With READ QUORUM/WRITE ANY satisfy the subSLA in $40$\% cases]{ \includegraphics[width=0.3\linewidth,height=4.2cm]{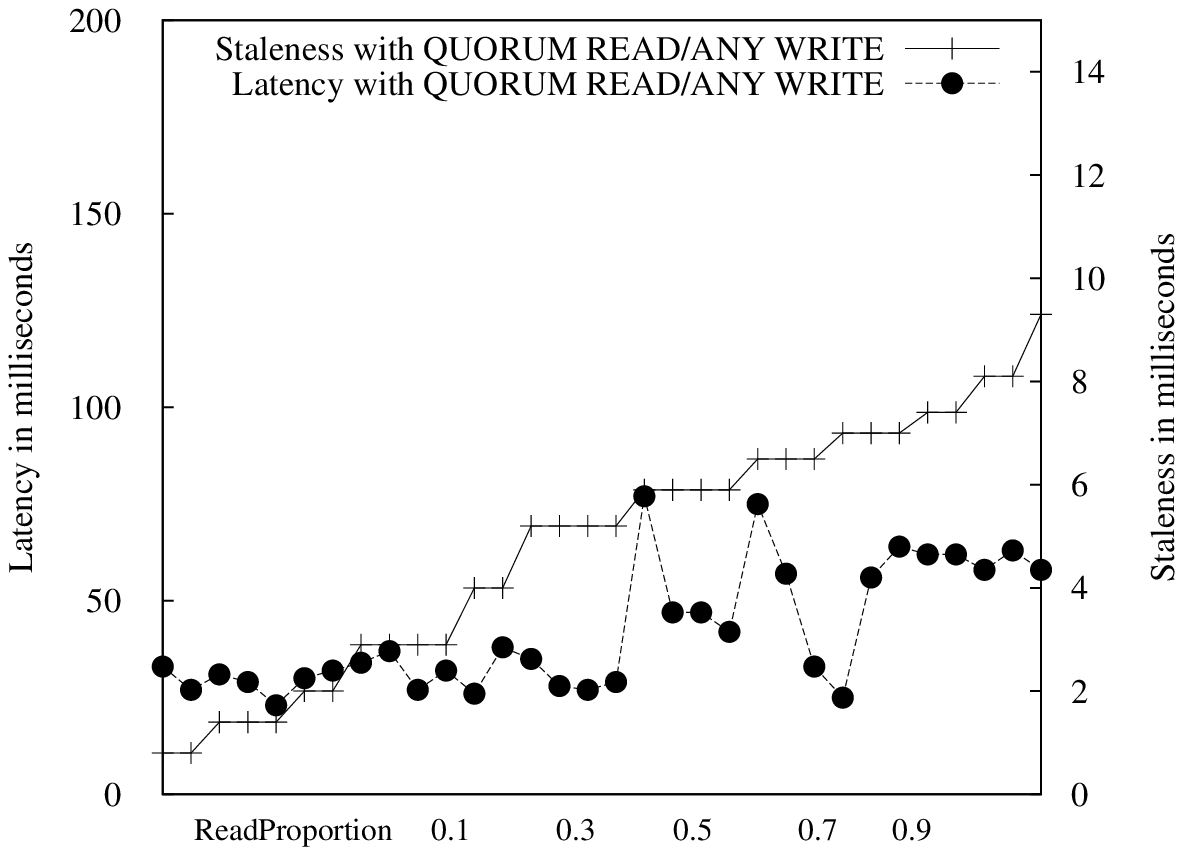}\label{fig:label-10}}\hfill
\subfigure[Operations With READ QUORUM/WRITE ONE satisfy the subSLA in $40$\% cases]{ \includegraphics[width=0.3\linewidth,height=4.2cm]{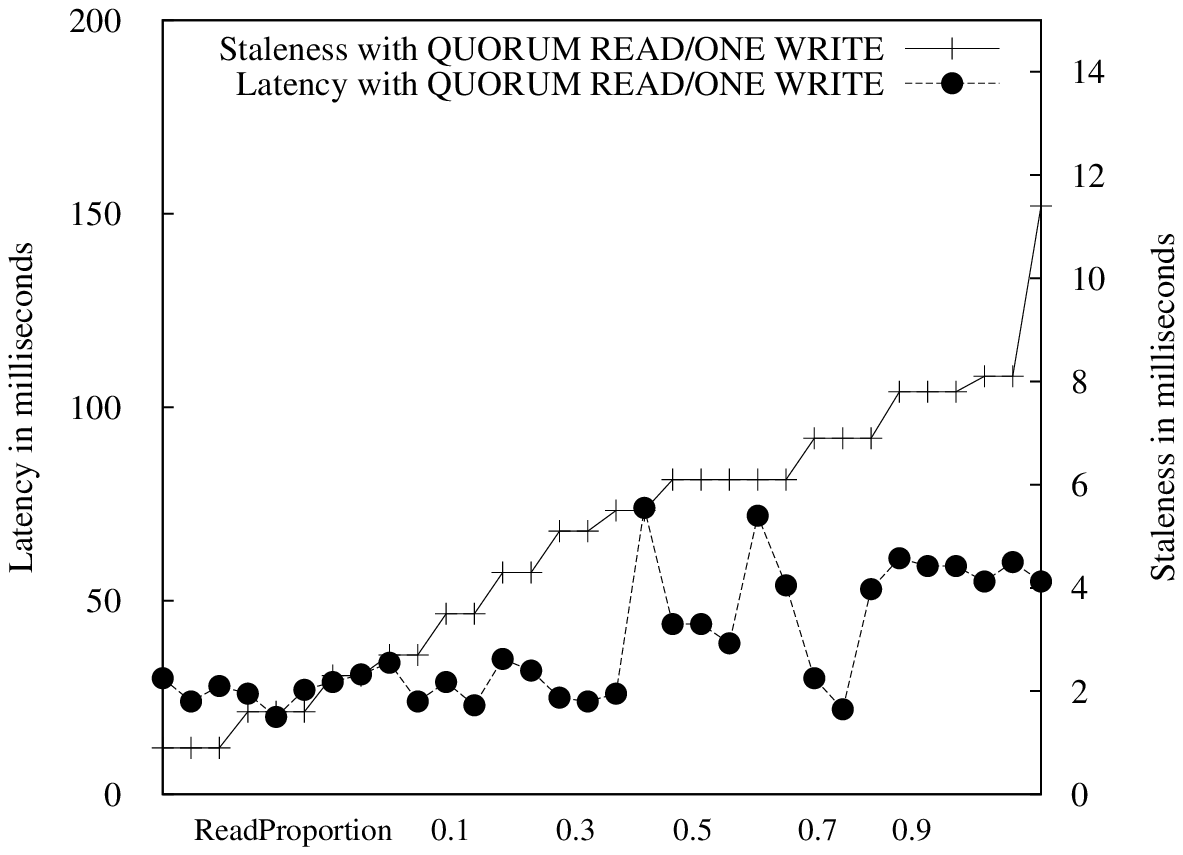}\label{fig:label-11}}\hfill
\subfigure[Operations With READ ONE/WRITE ANY satisfy the subSLA in $45$\% cases]{ \includegraphics[width=0.3\linewidth,height=4.2cm]{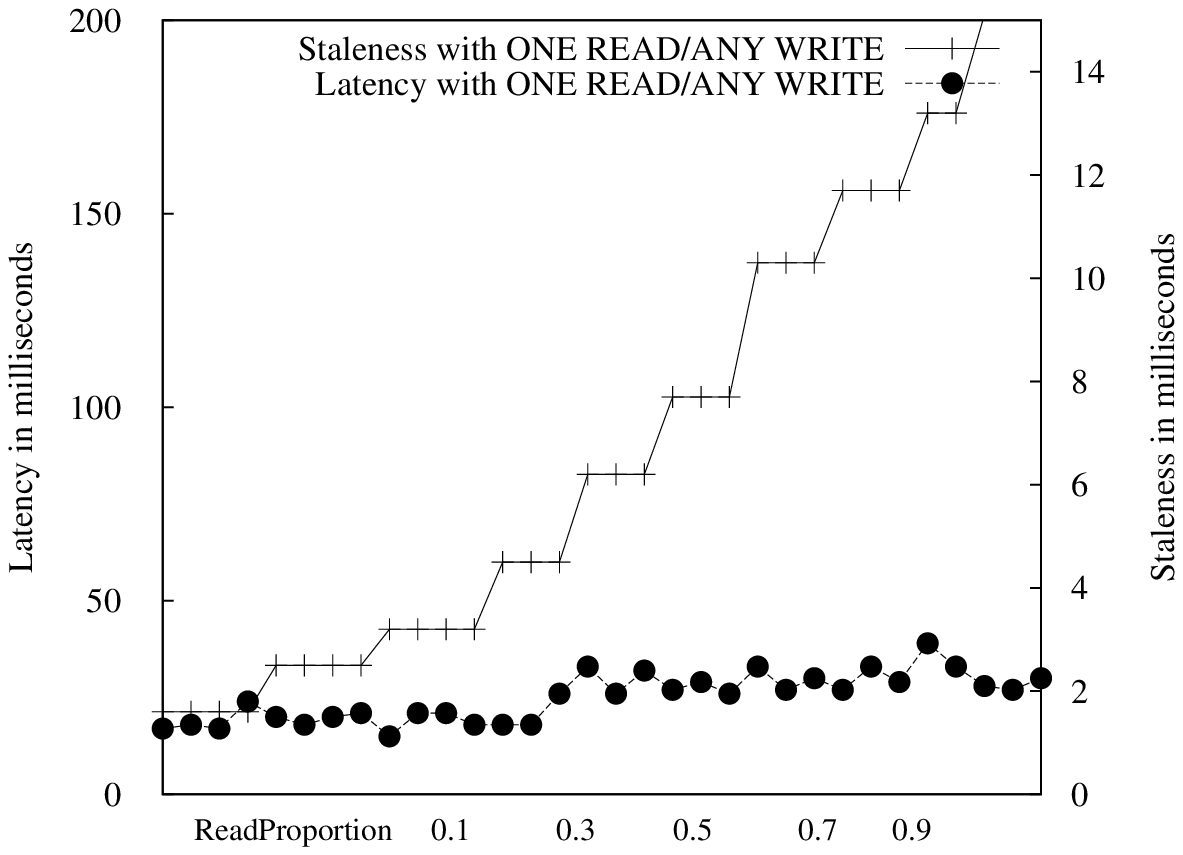}\label{fig:label-12}}\hfill
\subfigure[Operations With READ ONE/WRITE QUORUM satisfy the subSLA in $33$\% cases]{ \includegraphics[width=0.3\linewidth,height=4.2cm]{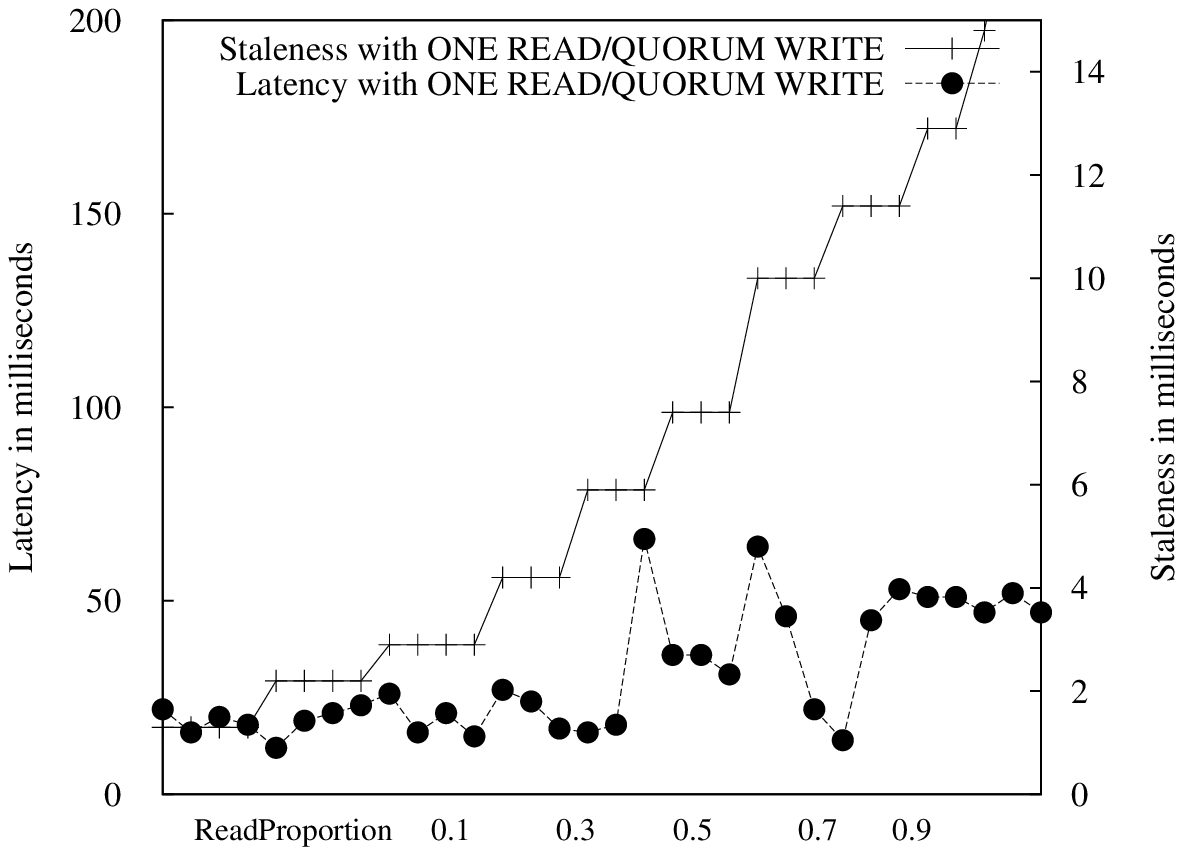}\label{fig:label-13}}\hfill
\subfigure[Operations With READ ONE/WRITE ALL satisfy the subSLA in $48$\% cases]{ \includegraphics[width=0.3\linewidth,height=4.2cm]{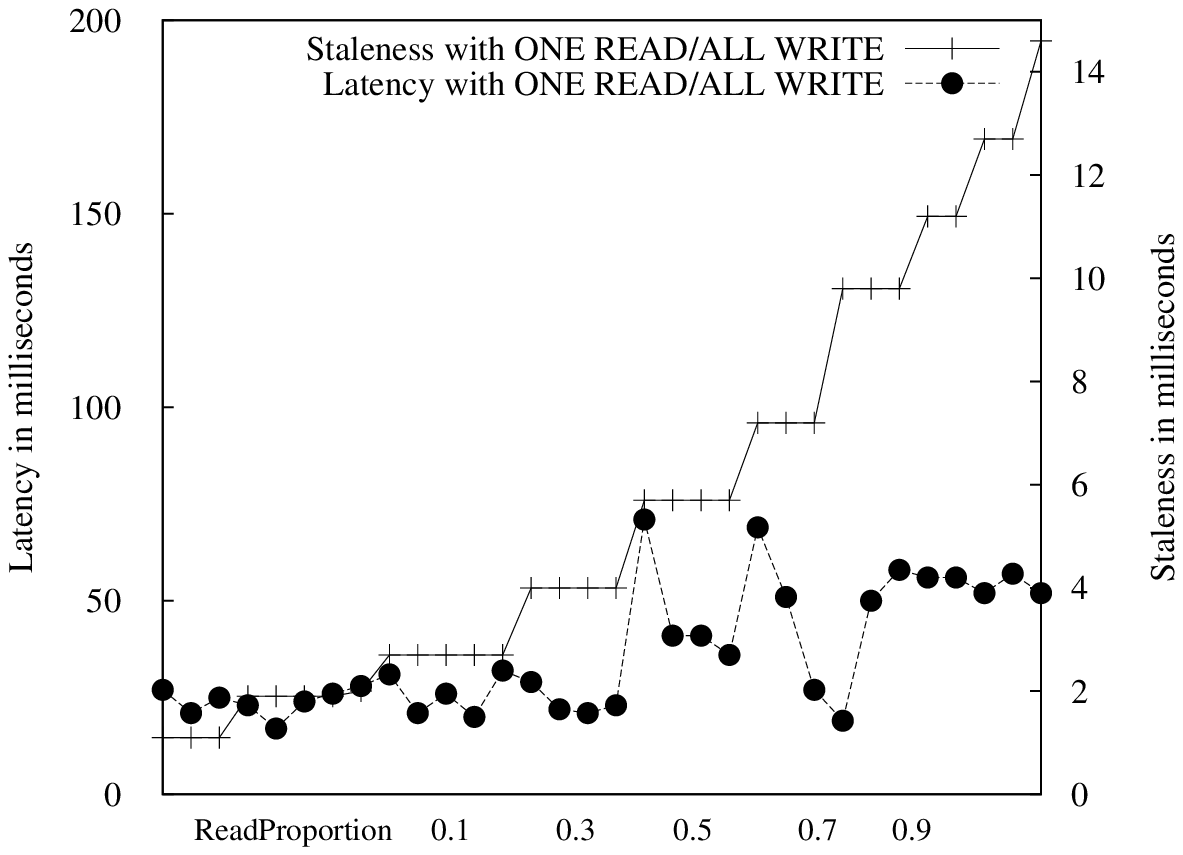}\label{fig:label-14}}\hfill
\caption{Adaptability of OptCon to Varying Workload (Read Proportion): Operations done with OptCon vs operations done with manually chosen consistency levels, under the subSLA SLA-1 (Latency:250ms Staleness: 5ms)}
%Manually chosen weak read-write consistency levels ONE/ANY are optimal (i.e., satisfy the subSLA) for low read proportions (i.e, in $<=55$\% cases) (Figures \ref{fig:label-8}, \ref{fig:label-9},
%\ref{fig:label-10}, \ref{fig:label-11}, \ref{fig:label-12}, \ref{fig:label-13}, and \ref{fig:label-14}). Manually chosen strong read-write consistency levels
%satisfy the subSLA for high read proportions (i.e, in $<=75$\% cases) (Figures \ref{fig:label-4}, \ref{fig:label-5}, \ref{fig:label-6}, and
%\ref{fig:label-7}).
%OptCon %is at least as effective as the optimal fixed consistency settings
% satisfies the subSLA (Figure \ref{fig:label-15}) for all possible read proportions by choosing the optimal consistency levels for each given read proportion.}
\end{figure*}

\begin{figure*}[!htpb]
\centering
%\noindent
\subfigure[Operations With Consistency Levels Predicted by OptCon under subSLA-1: Latency:250ms Staleness: 5ms]{\includegraphics[width=0.33\linewidth,height=3.3cm]{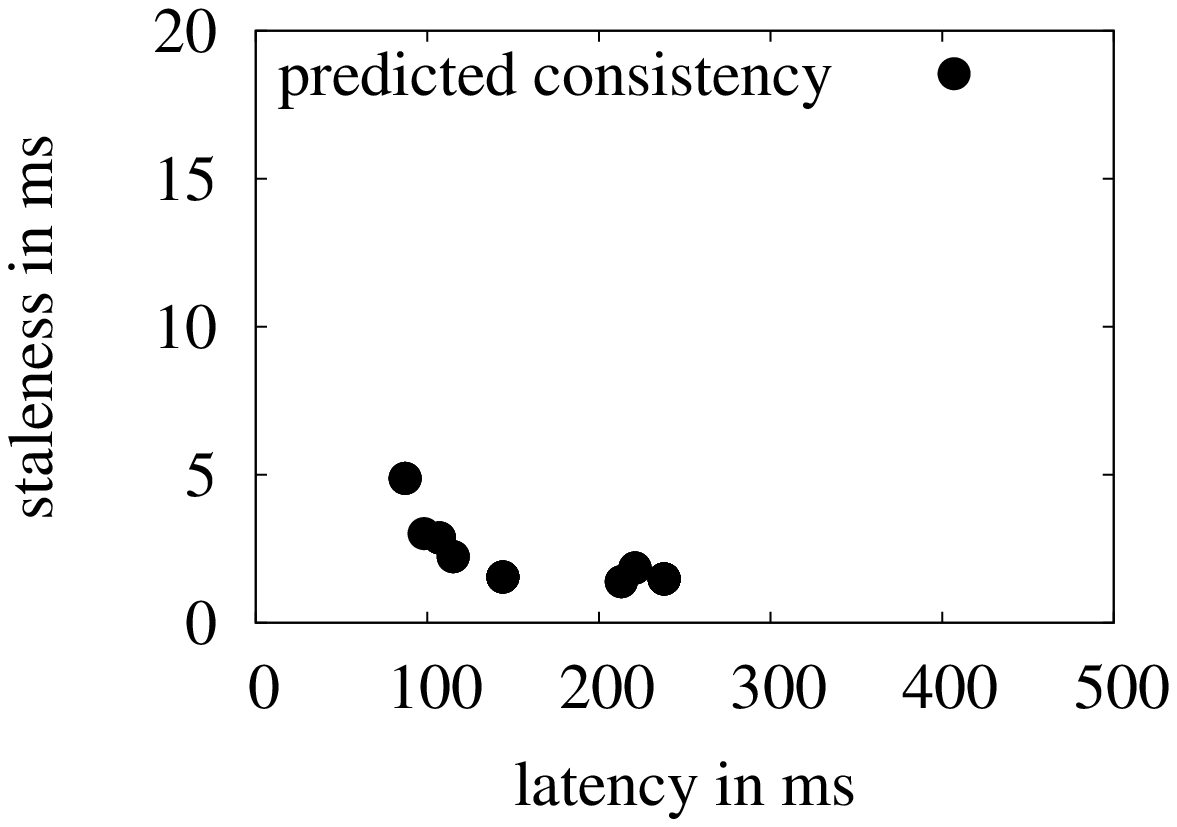}\label{label-a}}\hfill
\subfigure[Operations With Consistency Levels Predicted by OptCon under subSLA-2: Latency:100ms Staleness: 10ms]{\includegraphics[width=0.31\linewidth,height=3.3cm]{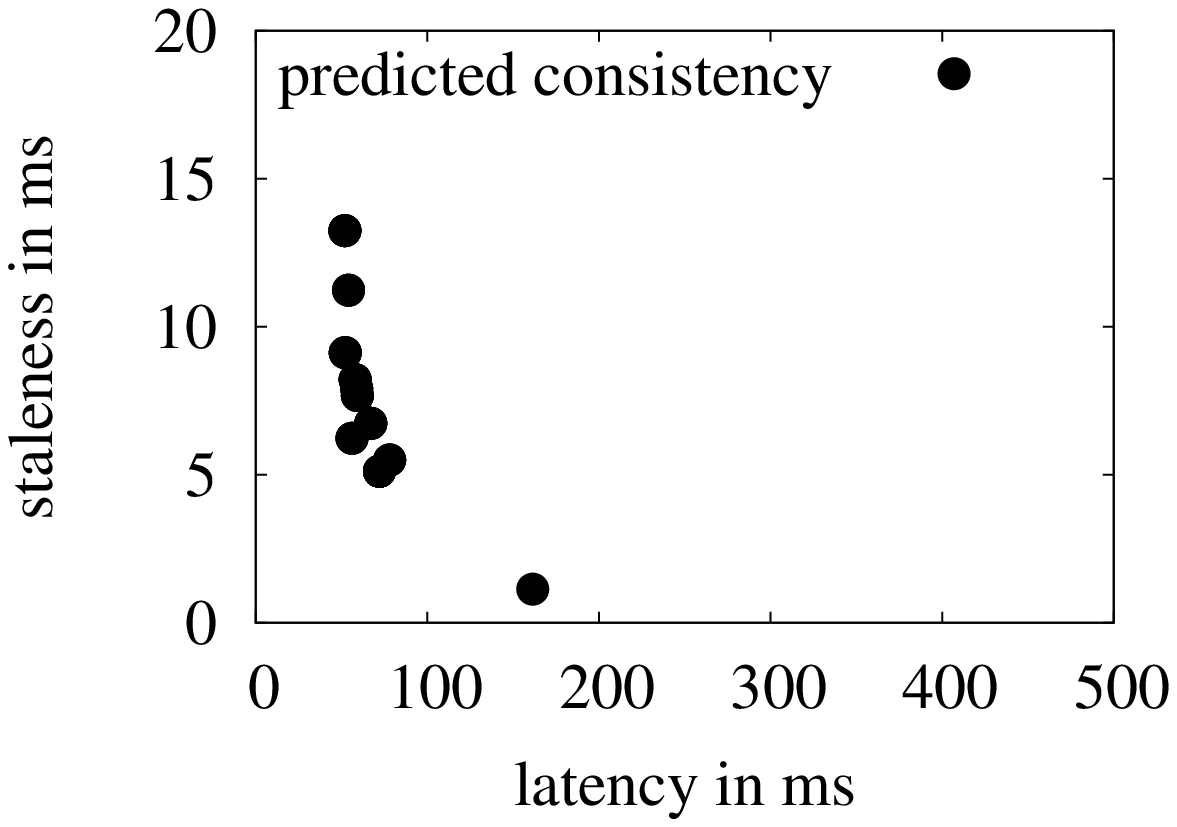}\label{label-b}}\hfill
\subfigure[Operations With Consistency Levels Predicted by OptCon under subSLA-3: Latency:20ms Staleness: 20ms]{ \includegraphics[width=0.31\linewidth,height=3.3cm]{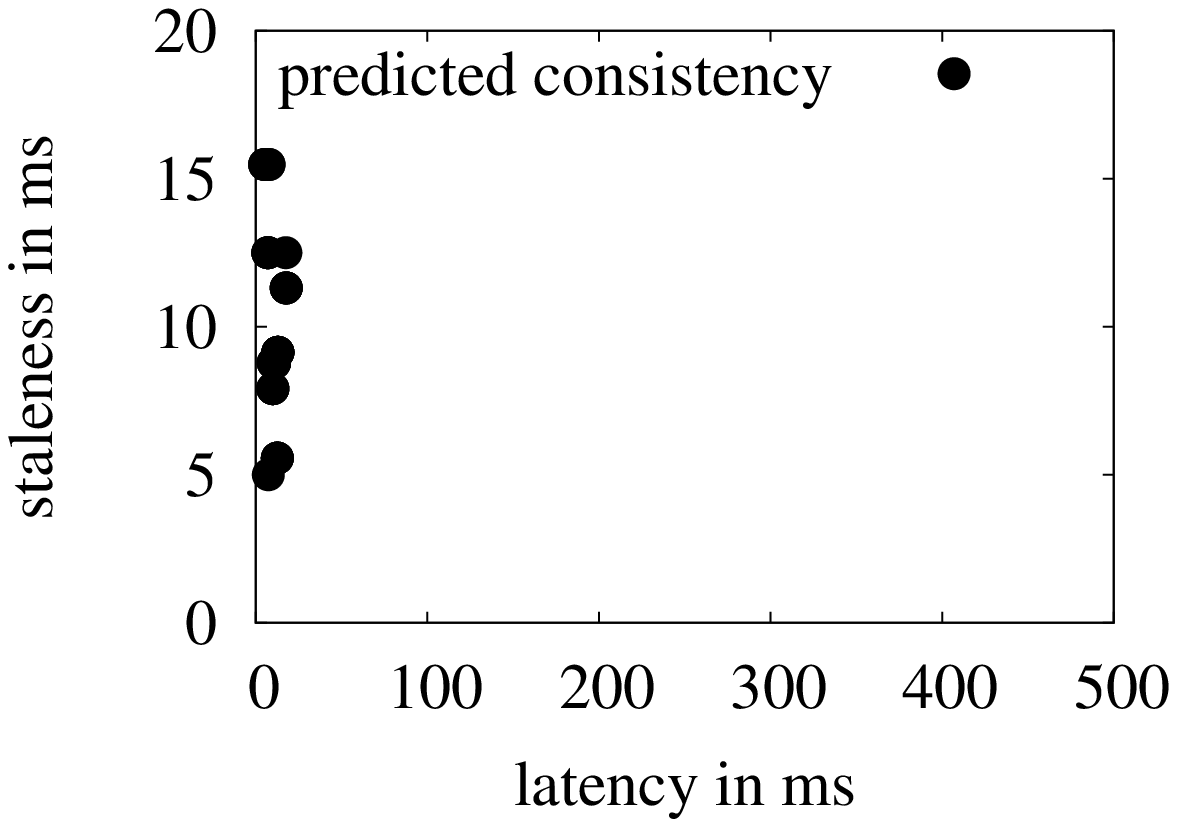}\label{label-c}}\hfill
\subfigure[Operations With READ ALL/WRITE ALL]{\includegraphics[width=0.31\linewidth,height=3.3cm]{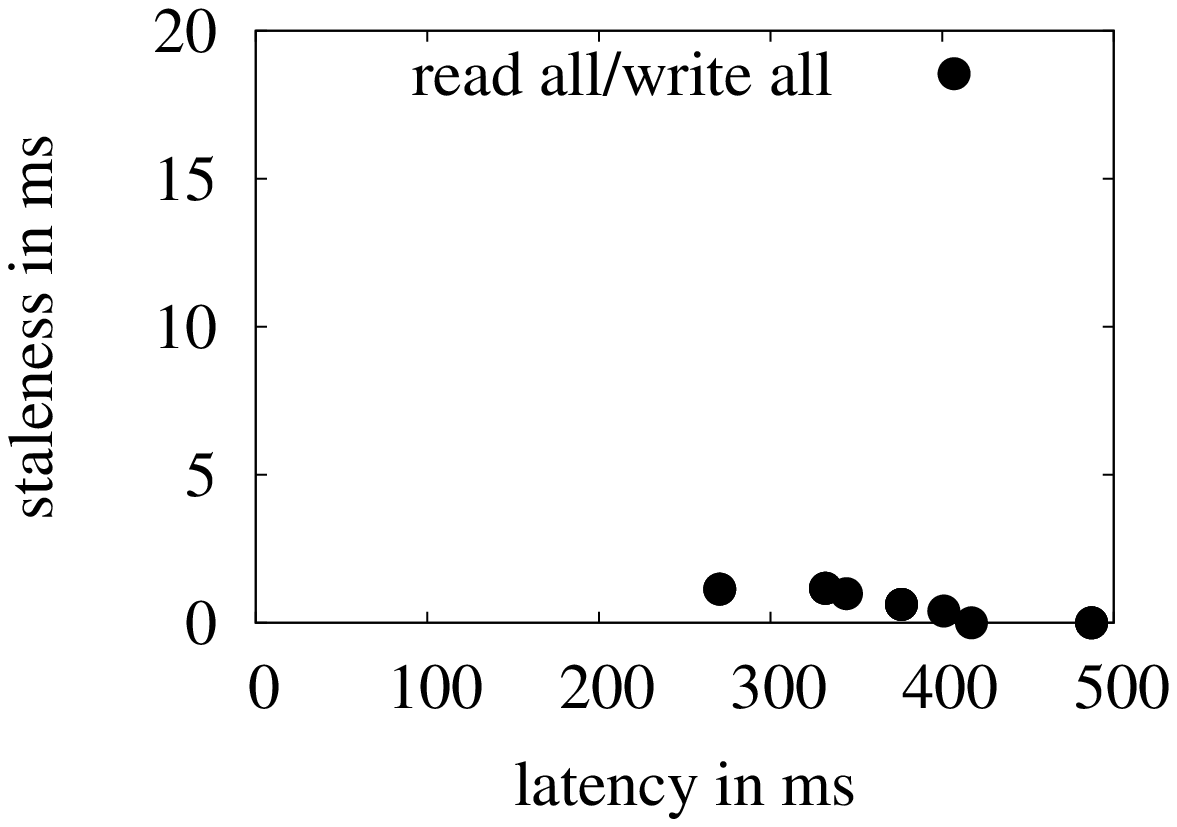}\label{fig:label-d}}\hfill
\subfigure[Operations With READ ALL/WRITE QUORUM]{\includegraphics[width=0.31\linewidth,height=3.3cm]{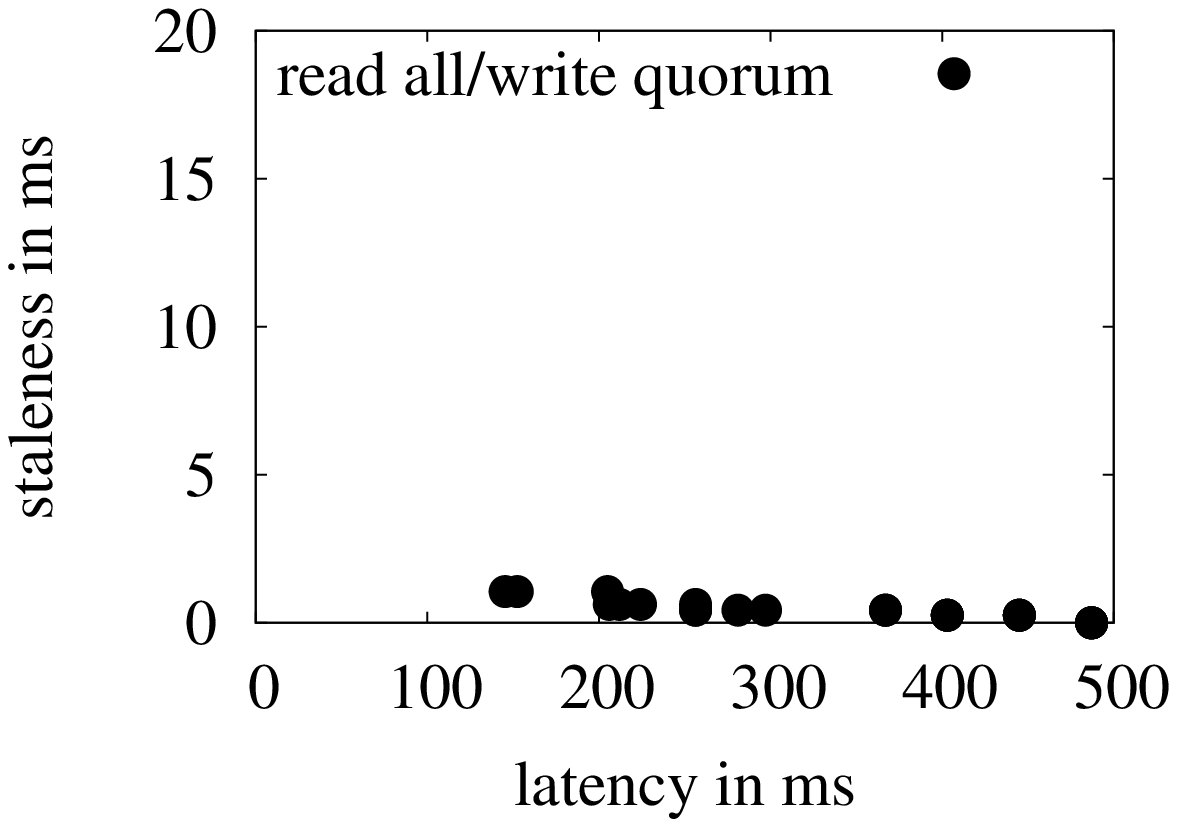}\label{fig:label-e}}\hfill
\subfigure[Operations With READ QUORUM/WRITE QUORUM]{ \includegraphics[width=0.31\linewidth,height=3.3cm]{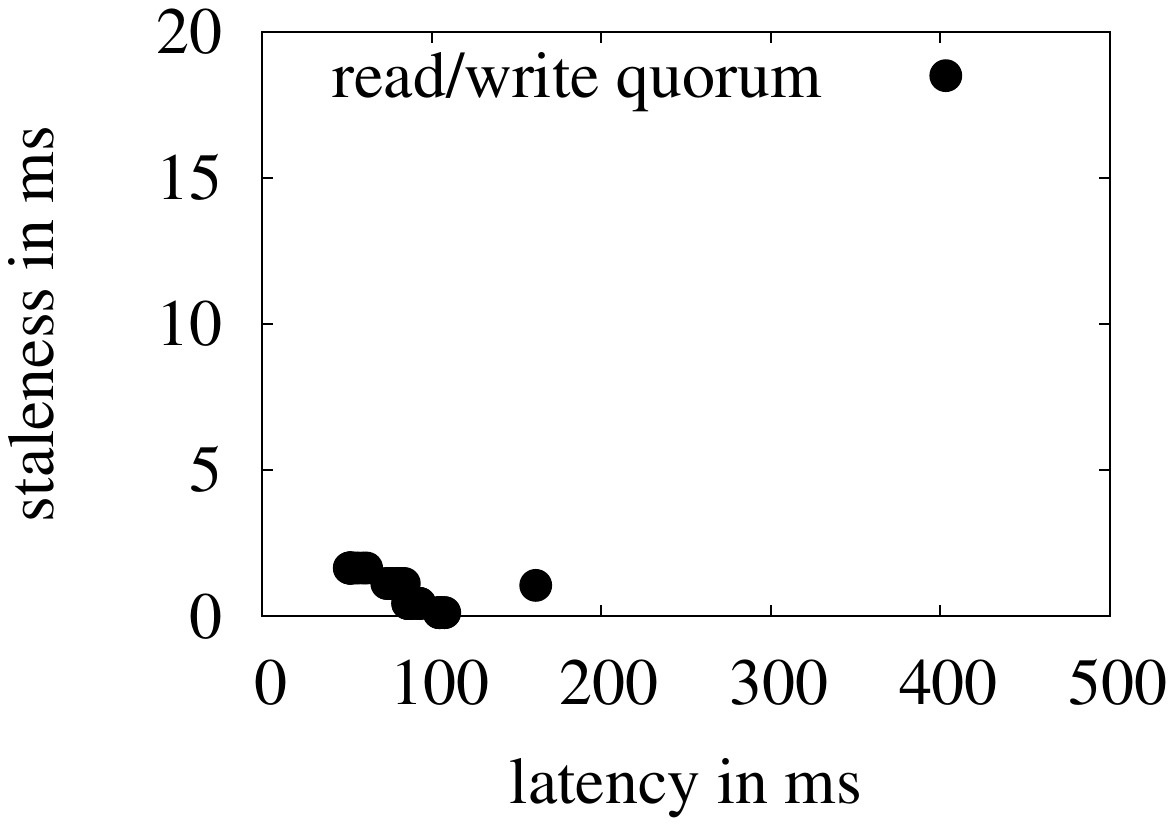}\label{fig:label-f}}\hfill
\subfigure[Operations With READ QUORUM/WRITE ALL]{ \includegraphics[width=0.31\linewidth,height=3.3cm]{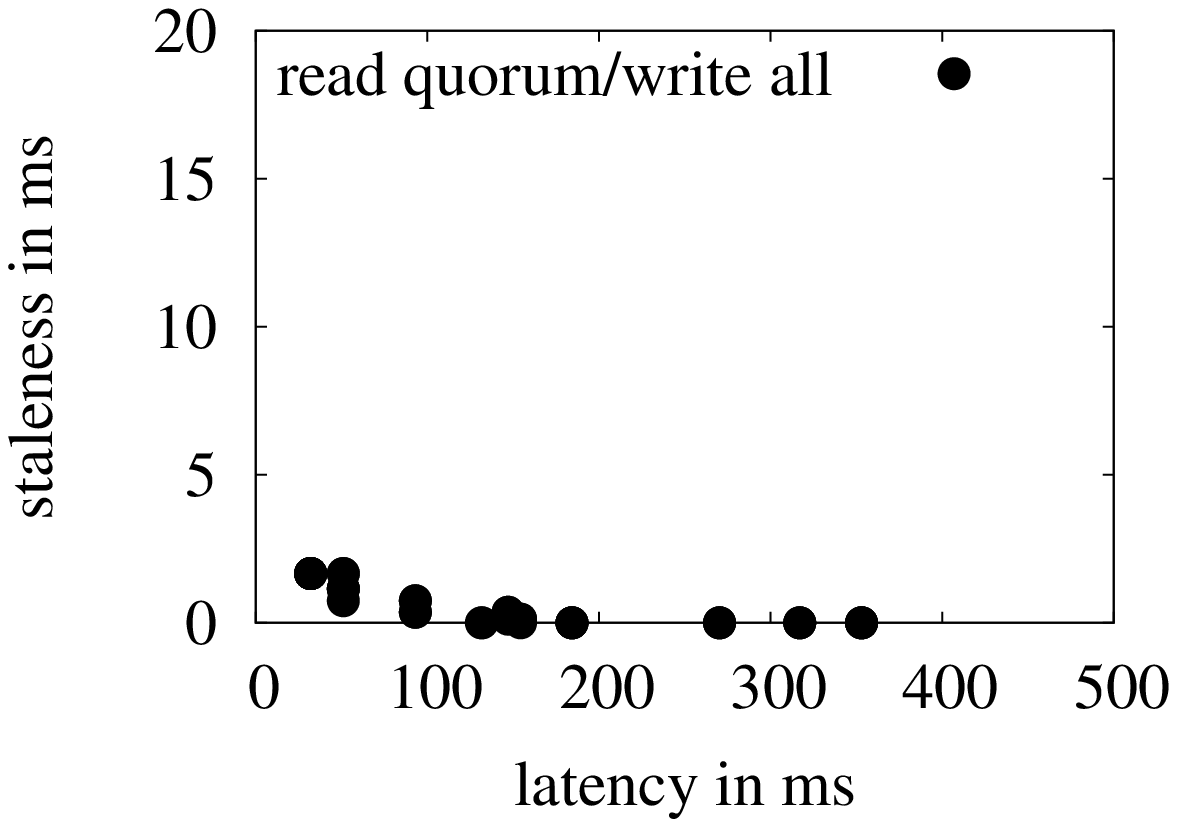}\label{fig:label-g}}\hfill
\subfigure[Operations With READ ALL/WRITE ANY]{ \includegraphics[width=0.31\linewidth,height=3.3cm]{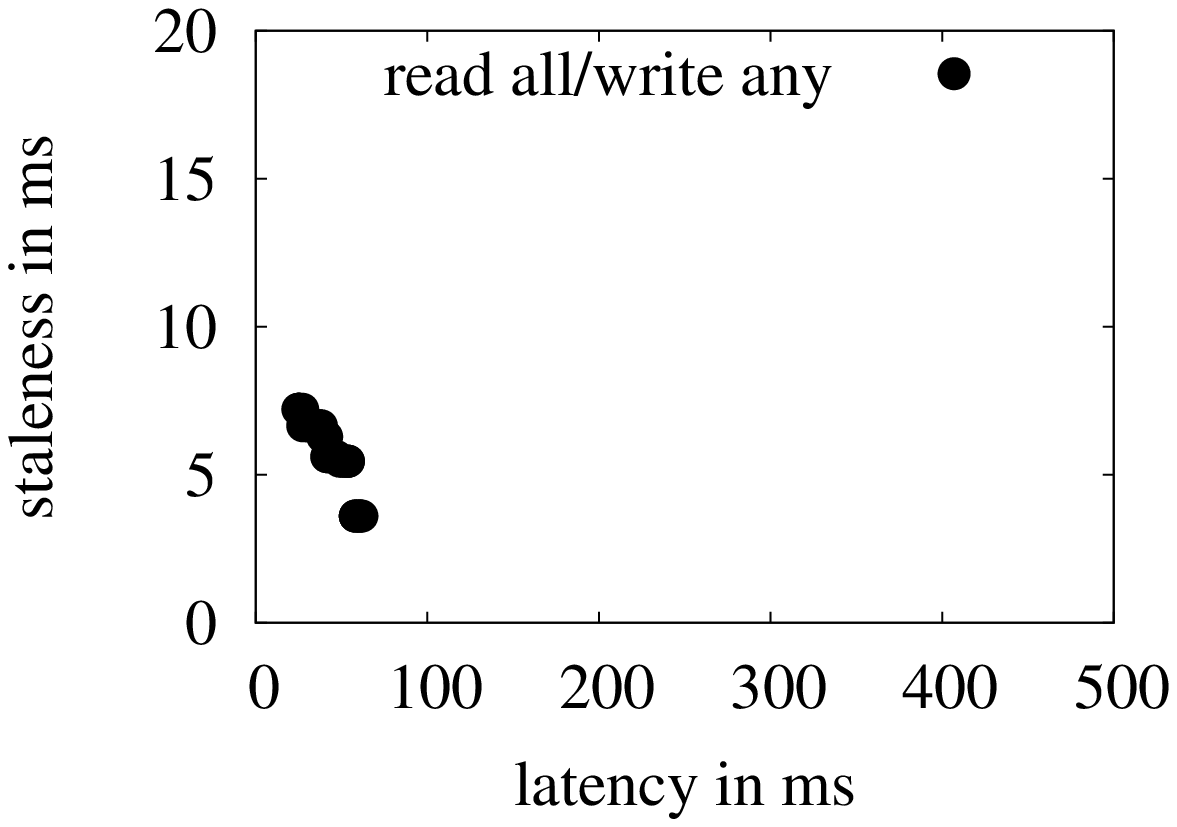}\label{fig:label-h}}\hfill
\subfigure[Operations With READ ALL/WRITE ONE]{ \includegraphics[width=0.31\linewidth,height=3.3cm]{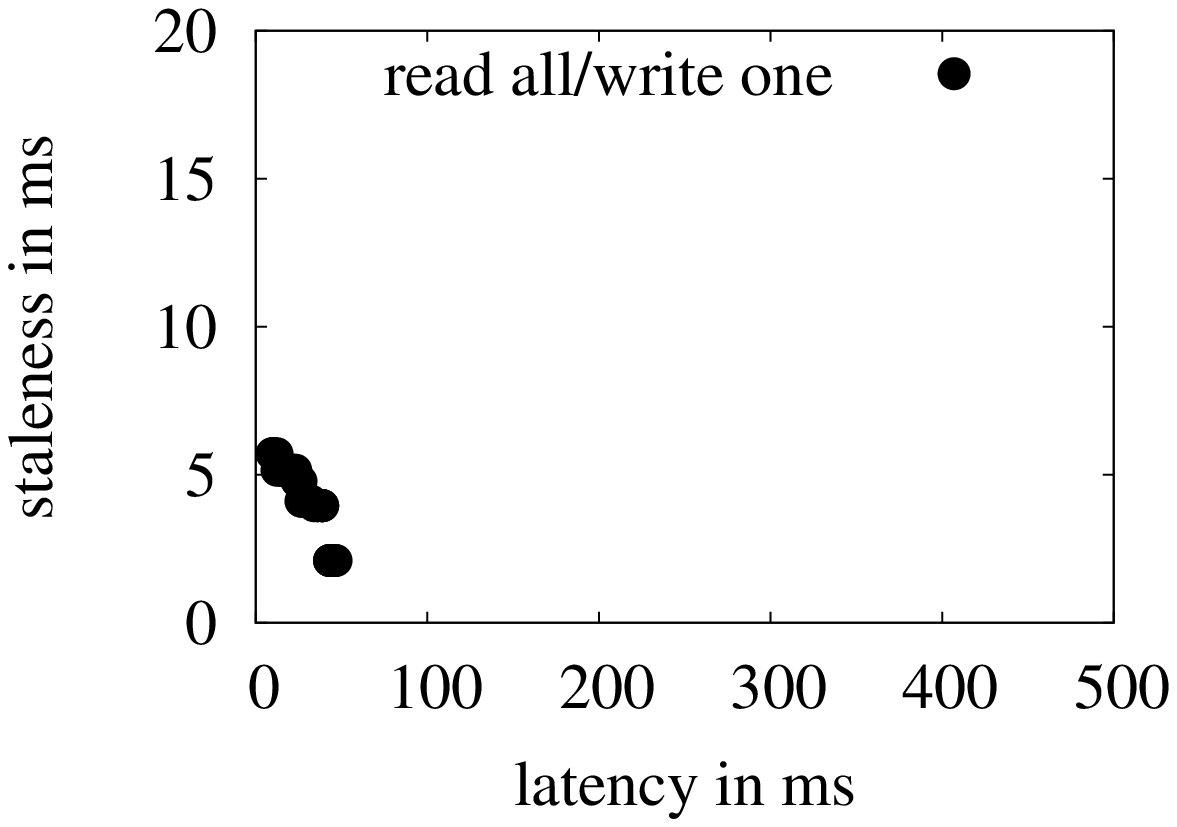}\label{fig:label-i}}\hfill
\subfigure[Operations With READ ONE/WRITE ALL]{ \includegraphics[width=0.31\linewidth,height=3.3cm]{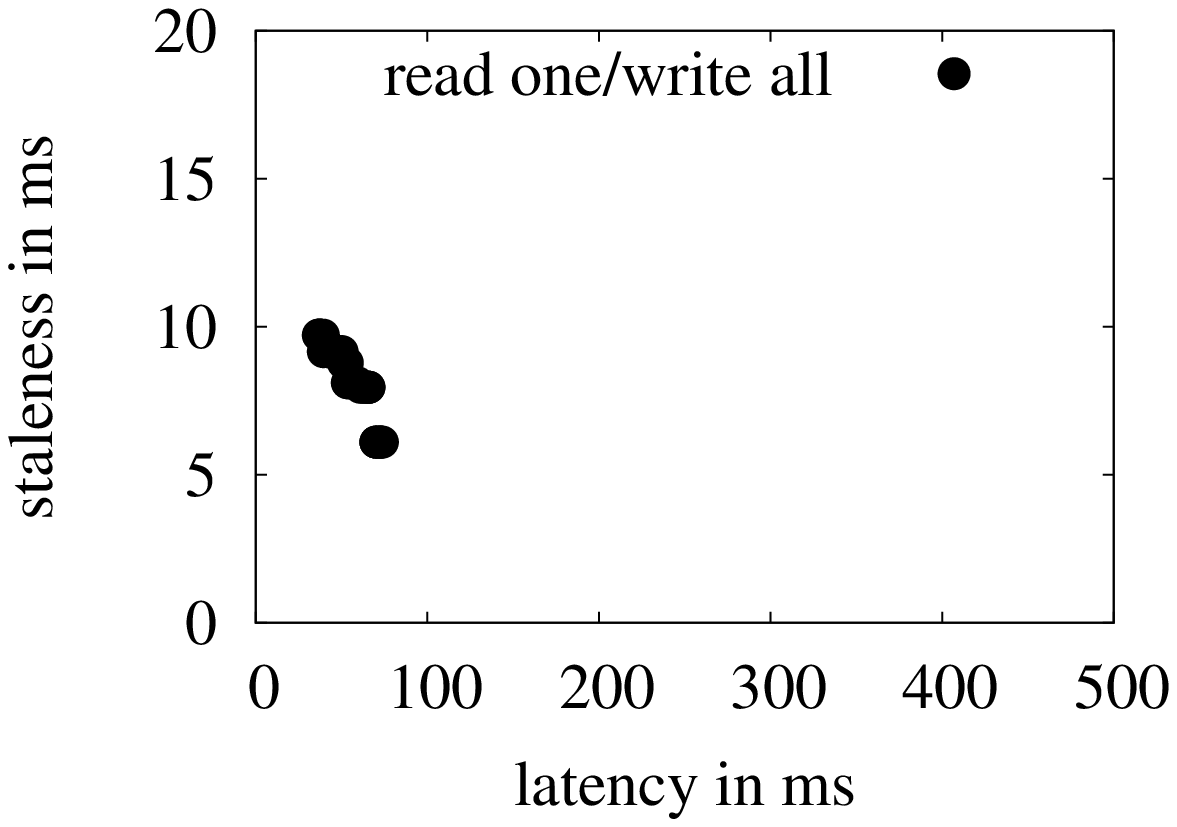}\label{fig:label-n}}\hfill
\subfigure[Operations With READ QUORUM/WRITE ONE]{ \includegraphics[width=0.31\linewidth,height=3.3cm]{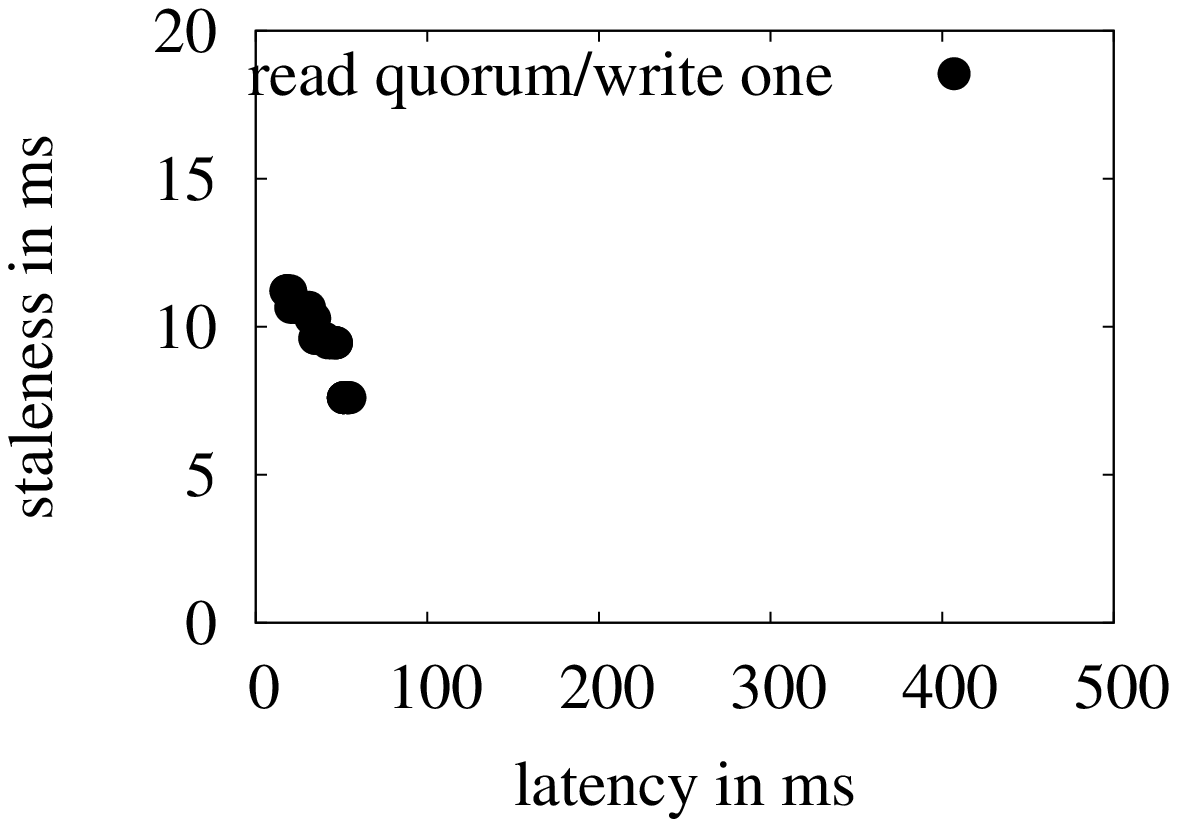}\label{fig:label-k}}\hfill
\subfigure[Operations With READ QUORUM/WRITE ANY]{ \includegraphics[width=0.31\linewidth,height=3.3cm]{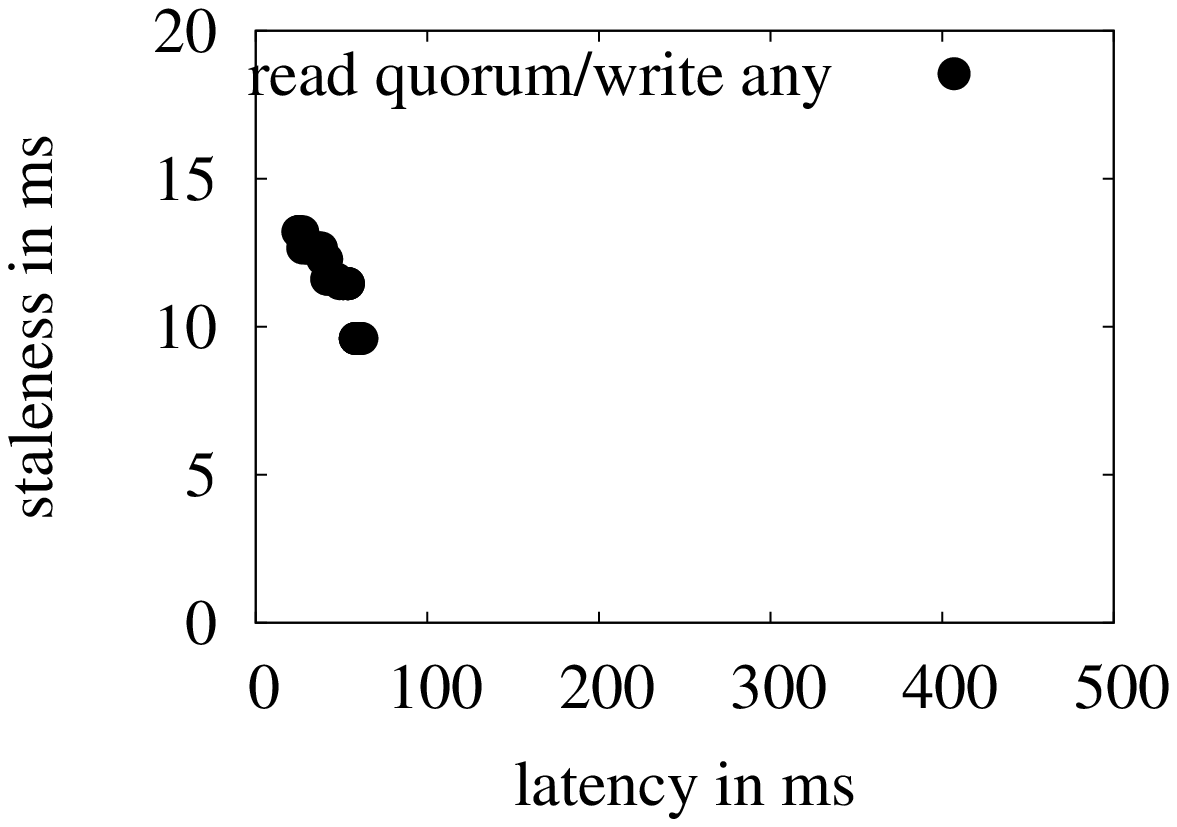}\label{fig:label-j}}\hfill
\subfigure[Operations With READ ONE/WRITE QUORUM]{ \includegraphics[width=0.31\linewidth,height=3.3cm]{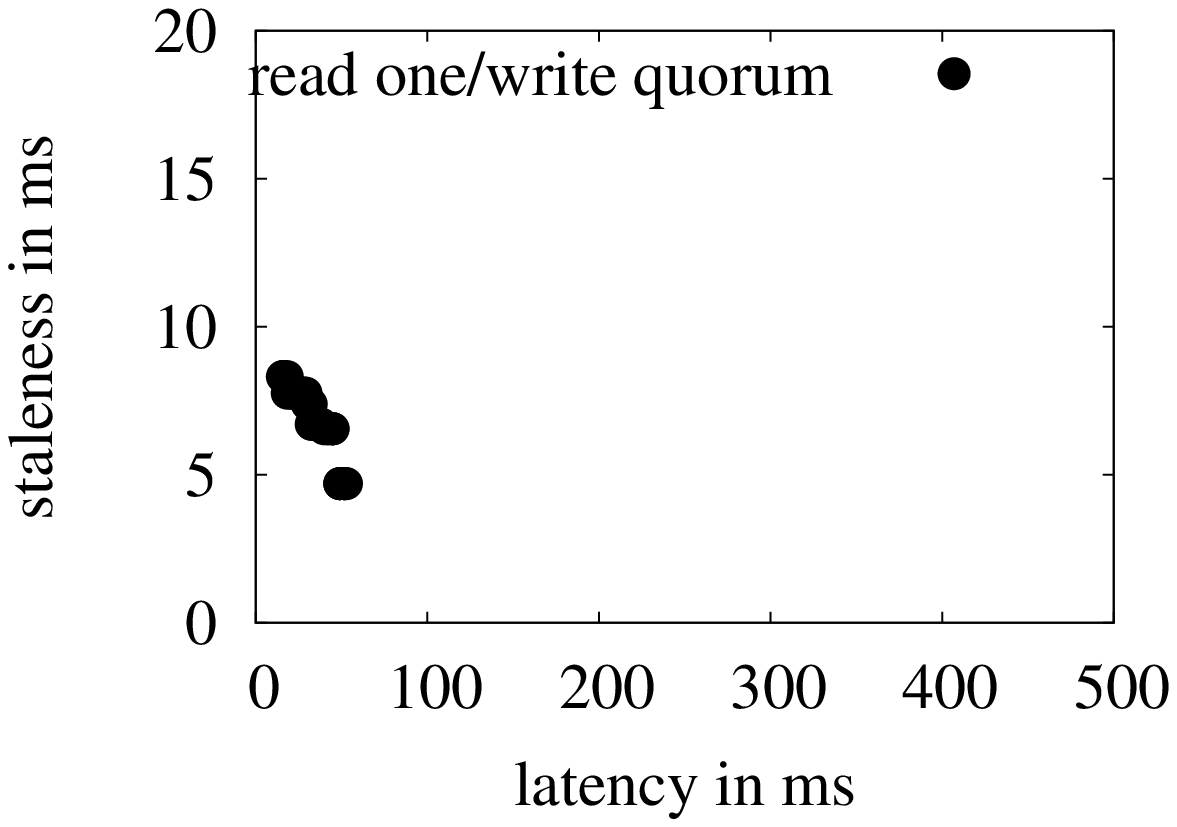}\label{fig:label-m}}\hfill
\subfigure[Operations With READ ONE/WRITE ANY]{ \includegraphics[width=0.31\linewidth,height=3.3cm]{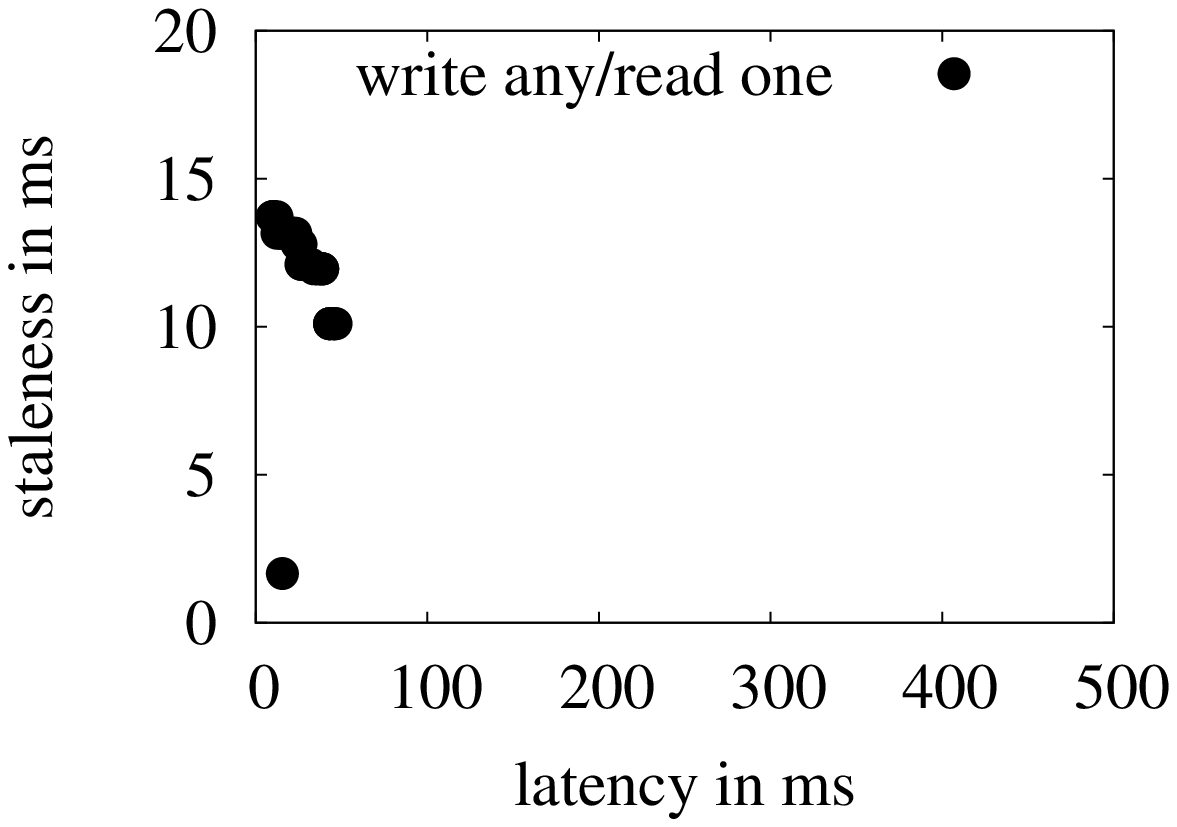}\label{fig:label-l}}\hfill
\subfigure[Chart Showing M-statistic values]{
\scalebox{0.8}{ \small\begin{tabular}[b]{|l|l|}
\hline
\textbf{Consistency Level}  & \textbf{M Value} \\ \hline
\multicolumn{2}{|l|}{\textbf{subSLA-1}}                              \\ \hline
WRITE ANY/READ ONE    & 68               \\ \hline
READ/WRITE QUORUM     & 41               \\ \hline
Predicted Consistency & 85               \\ \hline
\multicolumn{2}{|l|}{\textbf{subSLA-2}}                              \\ \hline
%\textbf{Consistency}  & \textbf{M Value} \\ \hline
READ QUORUM/WRITE ALL & 70               \\ \hline
Predicted Consistency & 92               \\ \hline
\multicolumn{2}{|l|}{\textbf{subSLA-3}}                              \\ \hline
%\textbf{Consistency}  & \textbf{M Value} \\ \hline
READ ALL/WRITE QUORUM & 0                \\ \hline
Predicted Consistency & 100              \\ \hline
\end{tabular}\label{table:mc}}}\hfill
\caption{Adaptability of OptCon to Different subSLAs: Operations done with OptCon vs operations done with manually chosen consistency levels}
%Manually chosen strong consistency
%levels (ALL) are optimal for SLA-1 with stringent staleness threshold
%(Figures \ref{fig:label-d}, \ref{fig:label-e}, \ref{fig:label-f}, and \ref{fig:label-g}). Manually chosen weak consistency
%levels (ANY/ONE) are optimal for SLA-3 with lower latency threshold (Figures \ref{fig:label-k},
%\ref{fig:label-j}, \ref{fig:label-m}, and \ref{fig:label-l}). Moderate consistency settings are
%sufficient for SLA-2.
%OptCon is as effective as the optimal consistency level for each subSLA (see Figures \ref{label-a}, \ref{label-b}, and \ref{label-c}).
 %Table \ref{table:mc} shows that OptCon (Figures \ref{label-a}, \ref{label-b}, and \ref{label-c}) is at least as effective as the optimal manually chosen consistency levels for each subSLA.} %OptCon only fails under the boundary cases caused by arbitrarily heavy network congestion.}
\end{figure*}

%We compare the predictive power of the different learning techniques used to learn the model. %Our analysis can provide the developers with the necessary guidance in making informed choice regarding the suitable learning technique.
 %Next in Section \ref{sec:metricsused}, we introduce the model selection metrics used. Section \ref{sec:compareresults} presents the results of model selection.
\subsubsection{The Model Selection Metrics Used}\label{sec:metricsused}
%Here we discuss the various model selection metrics used for comparing the learning techniques applied.
Cross validation error (CV error) \cite{Flach:2012:MLA:2490546} is the most common and widely used model selection metric. %, as it is easy to comprehend and compute.
 It represents the accuracy of the model with respect to an unseen test dataset, which is different from the original dataset used to train the model.  Using 10-fold cross validation, we partition the dataset into 10 subsamples, and run 10 rounds of validation. In each round a different subsample is used as the test dataset, while we train the model with the rest of the dataset. We compute the average of the mean squared error values for all validation rounds to obtain the mean CV error (Table \ref{table:selection}).
  We also use the Akaike Information Criterion (AIC) %\cite{Flach:2012:MLA:2490546}
 \cite{burnham2002model} %\cite{Akaike1974}
  which quantifies the quality of the generated model in terms of the information loss during the  training phase. For obtaining AIC, we first compute the likelihood of each observed outcomes in the test dataset with respect to the model. We compute the maximum log likelihood $L$, i.e.,  the maximum of natural logarithms over the likelihood of each of the observed outcomes. %The AIC metric is given as $\mathit{AIC} = 2k - 2\ln L,$ where $k$ is the number of parameters in the model.
  AICC \cite{burnham2002model} (Table \ref{table:selection}) further improves upon AIC, penalizing overfitting with a correction term for finite sample size n. Thus $\mathit{AICC} = 2k - 2\ln L + \frac{2k(k+1)}{n-k-1},$ where $k$ is the number of model parameters.
  Apart from AICC, we also compute the Bayesian Information Criterion (BIC) \cite{burnham2002model} %\cite{citeulike:90008}
   that uses the Bayesian prior probability estimates to penalize overfitting. Thus, $\mathit{BIC} = 2 k\ln N  - 2\ln L,$ % where $N$ is the number of examples in the training data.
    where the additional parameter $N$ for the sample size enables BIC to assign a greater penalty on the sample size than AICC. At smaller sample size, BIC puts lower penalty on the free parameters than AICC, whereas the penalty is higher for larger sample size (because it uses $k \ln N$ instead of $k$). %Table \ref{table:selection} gives the AICC and BIC metrics for each of the OptCon approaches.
   Normally, the BIC and AICC scores (Table \ref{table:selection}) are used together to compare the models, AICC detects the problem of overfitting and BIC indicates underfitting. Another criterion for the choice of the algorithm is the average prediction overhead (refer Section \ref{sec:over}), which is given in the last column of Table \ref{table:selection}.
  \subsubsection{Insights: Evaluation of Learning Techniques With Respect to the Metrics}\label{sec:compareresults}
  \begin{table}[!htb]
   % \begin{table}
  % title of Table
%\centering % used for centering table
\scalebox{0.9}{
\begin{tabular}{|c|c|c|c|c|} % centered columns (4 columns)
\hline\hline %inserts double horizontal lines
\bf Approach & \bf Cross Validation Error & \bf AICC & \bf BIC & \bf Overhead (ms) \\ % inserts table heading
 \hline\hline
Decision Tree & $0.14$ & $10.73$ & 51.44 & 1 \\ \hline
%SVM & $0.44$ & $11.21$ & 51.93 & Low \\ \hline % [1ex] adds vertical space
Bayesian Learning & $0.57$ & $12.85$ & 53.57 & 1.2 \\ \hline % [1ex] adds vertical space
Logistic Regression & $1.98$ & $16.32$ & 57.07 & 0.7 \\ \hline % inserting body of the table
Random Forest & $0.14$ & $9.51$ & 50.24 & 1.3 \\  \hline% [1ex] adds vertical space
Neural Network & $0.059$ & $12.85$ & 63.54 & 1.5 \\ % [1ex] adds vertical space
\hline %inserts single line
\end{tabular}
}
% is used to refer this table in the text
 \caption{Model Selection Results: As per descending order of AICC and BIC, and ascending order of CV and Overhead}
%\end{table}
\label{table:selection}
\end{table}
  %\begin{table}[!htb]
%   % \begin{table}
%  % title of Table
%%\centering % used for centering table
%\scalebox{0.94}{
%\begin{tabular}{|c|c|c|c|c|} % centered columns (4 columns)
%\hline\hline %inserts double horizontal lines
%\bf Approach & \bf Cross Validation Error & \bf AICC & \bf BIC & \bf Overhead (ms) \\ % inserts table heading
% \hline\hline
%Decision Tree & $0.14$ & $10.73$ & 51.44 & 1 \\ \hline
%%SVM & $0.44$ & $11.21$ & 51.93 & Low \\ \hline % [1ex] adds vertical space
%Bayesian Learning & $0.57$ & $12.85$ & 53.57 & 1.2 \\ \hline % [1ex] adds vertical space
%Logistic Regression & $1.98$ & $16.32$ & 57.07 & 0.7 \\ \hline % inserting body of the table
%Random Forest & $0.14$ & $9.51$ & 50.24 & 1.3 \\  \hline% [1ex] adds vertical space
%Neural Network & $0.059$ & $12.85$ & 63.54 & 1.5 \\ % [1ex] adds vertical space
%\hline %inserts single line
%\end{tabular}
%}
%% is used to refer this table in the text
% \caption{Model Selection Results}
%%\end{table}
%\label{table:selection}
%\end{table}
Table \ref{table:selection} gives the CV error, the AICC score, the BIC score, and the average prediction overhead, for each of the learning algorithms that OptCon has used.
   The above table can guide the developers in making an informed choice regarding the correct learning technique to use. The following analysis with respect to the Table \ref{table:selection} can act as lessons learnt for future practitioners and researchers looking to apply learning techniques to solve similar problems.
    Our problem can be directly cast as classification of the given training data into classes labelled (Section \ref{sec:approach}) by the consistency levels. % such that: 1) the SLA parameters staleness and latency satisfy the subSLA thresholds under given values of $RW$, $Tc$, and $P$, and subsequently 2) the secondary optimization criterion, namely throughput, is maximized.
    Hence, Decision Tree yields high accuracy and speed (Table \ref{table:selection}). We observed near random predictions with Linear regression (hence we omitted the results),  because: 1) our problem is more of a classification problem, as already explained, %, as explained in the next paragraph,
    and 2) linear regression requires the assumption of a linear model. %\cite{Flach:2012:MLA:2490546}.
     Logistic regression fairs worst among the approaches as indicated by the values of the metrics. This is because it also treats the problem as a  regression problem,  %\cite{Andrew:2007:STL:1273496.1273501}
  %and computes a smooth separation plane,
   whereas ours is a nonlinear classification problem. %(linear regression failed, and ours is a classification problem, as already elaborated).
    But it is still useful, since it yields a simple relation which expresses the effect of the parameters $RW$, $Tc$, and $P$, on $L$ and $S$ %average latency and staleness
   for an operation, under different consistency levels. Also, it can be used to determine the strength of the relationship among the parameters. Thus it can act as a basis for complex modelling algorithms. %For eliminating overfitting, we use the Lasso method to perform $L^1$ regularization on the model. We introduce to the minimization function additional regularization weights multiplied by a tunable constant coefficient $\Lambda$, expressed as a $L^1$ norm term. Particularly we use $\alpha$ = 0.05 and $\Lambda$ = 25.
%   subject to a constraint on the L1-norm of the coefficients B
  %Decision Tree yields high accuracy and speed (refer to Table \ref{table:selection}). This is because our problem can be directly cast as classification of the given training data into classes labelled (refer to Section \ref{sec:approach}) by the consistency levels such that: 1) observed staleness and latency satisfy the subSLA thresholds under given values of $RW$, $Tc$, and $P$, and 2) throughput is maximized. %In our experiments, we apply a confidence threshold of 0.25 for pruning, and one seed for random data shuffling.
%  Based on the values of the test variables comprising the consistency level and the knob parameters, OptCon must choose to proceed along a branch represented by specific values of staleness and latency.
   Random forest further eliminates errors due to overfitting and noisy data by: 1) bootstrapping multiple learners,  %\cite{Flach:2012:MLA:2490546},
    and 2) using randomized feature selection. Hence, it produces the best accuracy. %For our experiments we bootstrap 100 trees to train the random forest model.
    However, %training overhead owing to iterative learning, and
     the prediction overhead due to model complexity, might increase the latency overhead beyond the thresholds of real-time subSLA.
   Similarly, though ANN yields high accuracy (Table \ref{table:selection}), the additional overhead due to the model complexity (it produces the most complex model) may overshoot the threshold for real-time use cases. Bayesian method requires several approximating assumptions which result in high error scores.   % because of the problem of using SVM for nonlinear multi class classification.
 % In the absence of a closed form mathematical model, we cannot define a suitable kernel function for SVM. We tried linear kernel which yielded unacceptable scores.
 \subsubsection{Selection of Algorithm Based on Tradeoff Between Accuracy And Speed}\label{sec:criteria}

 \begin{table}[!htb]
   % \begin{table}
  % title of Table
%\centering % used for centering table
\scalebox{0.9}{
\begin{tabular}{|c|c|c|c|l|}
\hline\hline
\textbf{Decision Tree} & \textbf{Bayesian Learning} & \textbf{Logistic Regression} & \textbf{Random Forest} & \textbf{ANN} \\ \hline\hline
0.62                   & 0.46                       & 0.27                         & 0.52                   & 0.49         \\ \hline
\end{tabular}
}
% is used to refer this table in the text
 \caption{Evaluation of the Accuracy-Speed Tradeoff} % in Learning Techniques With the Measure \emph{$\mathit{Perf}$}}
%\end{table}
\label{table:choice}
\end{table}
  We define a measure \emph{$\mathit{Perf}$} for selecting the learning technique that provides an optimal tradeoff between the accuracy and speed (Table \ref{table:selection}).
   We assign the overhead of the slowest learning technique (as per Table \ref{table:selection}) %(refer to the last column in Table \ref{table:selection})
  as the baseline overhead $O_{\mathit{base}}$. Since ANN has the maximum overhead (1.5 ms) as per Table \ref{table:selection},
  $O_{\mathit{base}}$ = 1.5.
 The speedup $\mathit{Speedup}$ obtained with a chosen learning algorithm relative to the slowest
  algorithm is estimated from the observed relative decrease in overhead for the chosen algorithm with respect to the baseline $O_{\mathit{base}}$ (i.e., for the slowest algorithm, $\mathit{Speedup}$= 0).
   Thus,
  $\mathit{Speedup}$  = $\frac{O_{\mathit{base}} - O}{O_{\mathit{base}}}$, where $O$ is the overhead of the chosen
  technique.
     We denote the prediction error for a given learning technique as $E$. We assign
   the error for the least accurate algorithm, as per Table \ref{table:selection}, as the baseline
   $E_{\mathit{base}}$. Logistic regression has the maximum CV error (1.98 as per Table \ref{table:selection}).
    Hence, $E_{\mathit{base}}$ = 1.98. The value of $E$ and $E_{\mathit{base}}$ depends on the choice of
   the error measure.  The developer can either use the CV error, or both AICC and
   BIC taken together. %\cite{Flach:2012:MLA:2490546}.
    In the latter case, $E$ is given as the maximum between
   the AICC and BIC for the chosen algorithm, and $E_{\mathit{base}}$ is given as the maximum between the AICC and BIC
   for the least accurate algorithm.
   %If AICC and BIC are chosen, $E$ and $E_{\mathit{base}}$ are computed as the maximum
%   of the AICC and BIC scores for the respective algorithms.
    We compute the relative increase in accuracy ($A_{\mathit{Rel}}$) obtained with a chosen learning algorithm with respect
    to the least accurate algorithm. $A_{\mathit{Rel}}$ is estimated from the proportion of observed decrease in error for the chosen algorithm with respect to the baseline
    $E_{\mathit{base}}$. Thus,
  $A_{\mathit{Rel}}$  = $\frac{E_{\mathit{base}} - E}{E_{\mathit{base}}}$.
   Finally, the measure  $\mathit{Perf}$ is computed as $\mathit{Perf}$ = $w_{\mathit{Speedup}} \times \mathit{Speedup}$ + $w_A \times A_{\mathit{Rel}}$,
   where $w_{\mathit{Speedup}}$ and $w_A$ are the weights
  (real numbers in the closed interval $\mathclose{[}0, 1\mathclose{]}$, such that $w_{\mathit{Speedup}} + w_A = 1$) assigned by the developers to quantify
   the relative importance of speed and  accuracy, respectively, for selecting a learning technique. The
   developer can assign a larger weight to accuracy (i.e., $w_A > w_{\mathit{Speedup}}$), if accuracy is more important than speedup (and vice versa)
   according to the use case. The developer chooses the learning algorithm that produces the
   maximum value of $\mathit{Perf}$, with a given choice of error measure.
   We provide the values of $\mathit{Perf}$ for various learning algorithms in Table \ref{table:choice}, with
    CV Error as the error measure, and equal weights assigned to accuracy and speedup (i.e., $w_A$ = $w_{\mathit{Speedup}}$ = 0.5).
    The value of $\mathit{Perf}$ is maximum for the decision tree algorithm, % (i.e., $\mathit{Choice}$ = 0.62),
    implying that it produces an optimal tradeoff of accuracy and speed according to our choice of error measure, and the assigned values for the weights $w_A$ and $w_{\mathit{Speedup}}$.
    In most cases, the difference in overhead between algorithms is negligibly small, hence the $Speedup$ is
    not a criterion of significance for most cases. Thus, generally developers would assign weights $w_A$ = 1
    and $w_{\mathit{Speedup}}$ = 0, choosing the algorithm with the highest accuracy. %, with the above configuration.
   %Hence, the next experiments use decision tree implementation for the Learner module.

\subsection{Adaptability  of OptCon to Varying Workload}\label{sec:vary}%: Evaluation with Varying YCSB  ReadProportion}\label{sec:vary}
We demonstrate the adaptability  of OptCon to a varying workload, by tuning the
read proportion (i.e., proportion of reads) parameter ($RW$) in the YCSB benchmark workloads (Figure \ref{fig:label-15}).
%We use Decision tree for the Learner module as it gives the best results with equal weight to accuracy and speed  (refer to Section \ref{sec:criteria}).
 As per Table \ref{table:choice}, we choose the Decision tree  implementation of the Learner module. We compare operations performed
with OptCon, with those performed with manually chosen consistency levels. In the Figures \ref{fig:label-15}
through \ref{fig:label-14}, read proportion is plotted along the x axis, the observed latency along the primary y
axis, and the staleness along the secondary y axis.
  We demonstrate that, for a specific read proportion, only
 a subset of all possible read-write consistency levels is \emph{optimal}, i.e., satisfies latency and
 staleness thresholds in a given subSLA. Varying the read proportion ($RW$) in the workload from 0.1 to 1, we
 observe that a manually chosen read-write consistency level is optimal for only a fraction of the total number of cases,
 under the subSLA SLA-1 (Table \ref{table:sla}). OptCon demonstrates its adaptability to varying read proportion, applying
 the optimal consistency level for each read proportion, thus satisfying the subSLA in $100\%$ cases. OptCon can adapt to changing read proportion due to the presence of the parameter $RW$ as a feature in the prediction model $\mathcal{M}$. %Hence, a manually chosen fixed consistency level may not be matching for all possible read proportions.
 %OptCon demonstrates adaptability to varying read proportion, always applying consistency levels from the optimal set of consistency levels for each read proportion (refer  Figure \ref{fig:label-15}).
 As shown by the results in \cite{DBLP:conf/cloud/GolabRAKWG13}, the frequency of stale results (i.e., higher
$\Gamma$ scores) increases with increasing read proportion in the workload. Hence with higher read proportion %, i.e.,
 %$RW >$ 0.5 in
  in the right-half of the x axis, stronger read consistency levels are required to return less stale results. Also, since the proportion
 of writes is less, the penalty for the write latency overhead is less. Hence stronger write consistency levels, that result in high write latency, are acceptable. Thus for higher read proportions, combinations of strong read-write consistency levels, i.e., ALL/QUORUM, succeed in lowering staleness values to acceptable bounds (right-half of the Figures \ref{fig:label-4}, \ref{fig:label-5}, \ref{fig:label-6},  and  \ref{fig:label-7}).  Strong consistency levels achieve a maximum success rate of 75\% in satisfying the subSLA, with READ ALL/QUORUM WRITE (Figure \ref{fig:label-5}) and READ QUORUM/WRITE ALL (Figure \ref{fig:label-6}. Taking into account the clock skew, the staleness is effectively 0 in these cases.  %But, weak read consistency settings result in high staleness values for  workloads with high read proportion.
   For the same reasons, for higher read proportions, weak read consistency levels result in failure to achieve stringent staleness thresholds, as demonstrated by the points in the right-half of
  Figures \ref{fig:label-8} through \ref{fig:label-14}. %Figures \ref{fig:label-4}, \ref{fig:label-5}, \ref{fig:label-6},   and
%  \ref{fig:label-7} demonstrate that the stronger manually chosen read-write consistency levels, i.e., ALL/QUORUM, succeed in lowering staleness values to acceptable bounds under high read proportion.
 With lower read proportions in the left-half of the x axes, the frequency of stale read results are smaller \cite{DBLP:conf/cloud/GolabRAKWG13}. In this case,
weaker manually chosen read-write consistency levels, i.e., ONE/ANY (left-half of Figures \ref{fig:label-8} through \ref{fig:label-14}) are sufficient for returning
consistent results. For lower read proportions, stronger consistency levels, i.e., ALL or QUORUM, are unnecessary,
only resulting in high latency overhead. With weak consistency levels, we observe a maximum success rate of only $<=55$\% in  satisfying the subSLA (left-half of Figures \ref{fig:label-4} through \ref{fig:label-7}), with READ ONE/WRITE ALL (Figure \ref{fig:label-14}). %Thus under lower read proportions, weaker manually chosen consistency levels, i.e., ONE/ANY, give better results, i.e., both lower staleness and lower latency, than ALL or QUORUM settings.
  % Thus, a particular manually chosen fixed consistency setting can satisfy staleness and latency thresholds in a given SLA for only certain workloads.
% Strong consistency levels are suitable for higher read proportions, while weaker consistency is sufficient for lower read proportions.
  Thus, manually chosen consistency levels show a maximum success rate of 75\% in satisfying the subSLA, % with READ ALL/QUORUM WRITE (Figure \ref{fig:label-5}) and READ QUORUM/WRITE ALL (Figure \ref{fig:label-6}.
 in contrast with a $100$\% success rate demonstrated by OptCon (Figure \ref{fig:label-15}).
Thus, OptCon is more effective than any manually chosen consistency level in adapting to workload variations.

\subsection{Adaptability  of OptCon to Different subSLAs}\label{sec:user}%: User Simulation with RUBBoS}\label{sec:user}
 %We demonstrate the adaptability  of OptCon to the latency and staleness thresholds of different subSLAs.
  For a given subSLA, only a subset of all possible manually chosen consistency levels is \emph{optimal}, i.e., satisfies the subSLA thresholds. Consistency levels predicted by OptCon are always matching, i.e., it always chooses from the above optimal set of consistency levels for the given subSLA. %But, a manually chosen fixed consistency setting may not always produce optimal results for a given subSLA.
  We have integrated OptCon (using Decision tree learning as per Table \ref{table:choice})
with the RUBBoS \cite{objWeb-RUBBoS} benchmark, that can simulate concurrent access, and interleaving user access. %We performed user simulation with standard operations like user registration, login, search, browse, submission, and review.
%     OptCon as a wrapper over RUBBoS.
 Figures \ref{label-a} through \ref{fig:label-l} plot the observed staleness along the y axis, and latency along the x axis, under subSLAs SLA-1, SLA-2 and SLA-3, respectively (Table \ref{table:sla}). Figures \ref{label-a} through \ref{label-c} demonstrate experiments performed with OptCon. Figures \ref{fig:label-d} through \ref{fig:label-n} correspond to experiments done with manually chosen consistency levels.
 We demonstrate a few extreme cases of occasional heavy network traffic in real
world applications with artificially introduced network delays (refer to the few boundary cases with arbitrarily high latency in Figures
\ref{label-a}, \ref{label-b}, and \ref{label-c}, that violate the respective subSLAs). These boundary cases were simulated using the traffic shaping feature of Traffic Control \cite{2004lartc_howto}, a linux-based
tool for configuring network traffic, that controls the transmission rate by lowering the network
 bandwidth.
 SLA-1 (Table \ref{table:sla}) represents systems demanding stronger consistency. %, such as transaction and batch processing applications.
 Manually chosen weak read-write consistency level, i.e, ANY/ONE, fails to satisfy the stringent staleness bound of SLA-1 (Figures \ref{fig:label-k} through \ref{fig:label-m})  %with fixed weak (eventual) read-write consistency settings, i.e, ANY/ONE, the system fails to satisfy the stringent staleness bound in SLA-1.
  On the other hand, strong consistency levels (i.e., ALL) (Figures
\ref{fig:label-d}  through \ref{fig:label-g}) satisfy SLA-1. %produce optimal results --- achieving
%low staleness, while compromising on the latency front.
 OptCon satisfies SLA-1 by applying the respective optimal consistency levels (i.e., ALL in this case) for SLA-1 (Figure \ref{label-a}). %--- thus matching the optimal consistency settings for SLA-1.
  For lower latency bound in SLA-3 (Table \ref{table:sla}), %represents situations where the use case demands weak consistency
%settings, like real time \cite{Terry:2013:CSL:2517349.2522731}
%systems. % (refer to Figure \ref{fig:usecase}).
  the manually chosen strong consistency levels (ALL/QUORUM) unnecessarily produce latencies beyond the acceptable threshold (Figures
\ref{fig:label-d}  through \ref{fig:label-g}). However, weaker consistency levels (i.e., ONE/ANY) prove to be optimal (Figures \ref{fig:label-k} through \ref{fig:label-n}). Again, OptCon  succeeds by choosing the respective optimal consistency levels (i.e., ONE/ANY in this case) for SLA-3 (Figure \ref{label-c}).
 Similarly with SLA-2, a subset of manually chosen fixed consistency levels (Figures \ref{fig:label-h}, \ref{fig:label-i}, and \ref{fig:label-n}) produce optimal results, whereas OptCon successfully achieves SLA-2 in Figure \ref{label-b} for all operations. %Thus OptCon adapts to varying subSLA demands, choosing the respective optimal consistency levels for each subSLA.
 We evaluate OptCon by a measure \emph{$M$}, which measures the adaptability %\cite{wiki:xxx}
 %\cite{paschke2006categorization}
  of the system with the subSLA. %, and 2) $M_c$, which measures the performance of the system with respect to the maximization objective. Inspired by the average utility metric used by \cite{Terry:2013:CSL:2517349.2522731}, we represent $M_c$ as a percentage improvement over fixed consistency level.
  Following \cite{Terry:2013:CSL:2517349.2522731}, $M$ (Table \ref{table:mc}) computes the percentage of cases which did not violate the subSLA. For all the given subSLAs, the $M$ values for operations performed with OptCon, given in the Figures \ref{label-a}, \ref{label-b}, and
\ref{label-c}, exceed the $M$ for operations using fixed consistency levels. Thus, OptCon is at least as effective as any optimal
  consistency level for the given subSLA. Moreover, OptCon produces matching
choices where fixed consistency levels fail. % to satisfy the subSLA demands. %The experiments represent the common operations for any desktop or web based applications.
 % Thus, OptCon can upgrade real world storage applications to work under varying subSLAs.

\section{Related Work}\label{sec:related}
%As it is impossible to achieve all the three objectives of
% Brewer's CAP principle %(i.e., consistency, availability, and partition-tolerance
% \cite{brew:cap,Gilbert:2002:BCF:564585.564601,Abadi:2012:CTM:2360751.2360959}),  distributed datastores %\cite{Lakshman:2010:CDS:1773912.1773922, Sumbaly_servinglarge-scale, Meiklejohn:2013:RPD:2505305.2505309, Plugge:2010:DGM:1869938, Schutt:2008:SRT:1411273.1411280, DeCandia:2007:DAH:1323293.1294281}
% overcome this limitation by offering relaxed consistency guarantees \cite{birman1996trading} to provide high availability.
Wada et al. \cite{wada:cidr} and Bermbach et al. \cite{Bermbach} analyze and quantify %system-centric
consistency % in commercial cloud storage systems %,
%	and answered the question ``how soon is eventual?''
 from a system-centric perspective that
	focuses on the convergence time of the replication protocol.
Bailis et al. \cite{Bailis:2012:PBS:2212351.2212359} and Rahman et al. \cite{Rahman:2012:TPF:2387858.2387866} instead consider a client-centric perspective of consistency,  in which a consistency anomaly occurs
	only if differences in state among replicas lead to a client application actually observing a stale read.
        In practice eventual consistency is preferred over strong consistency in scenarios where the system must maintain availability during network partitions \cite{birman1996trading, bayou},
	or when the application is latency-sensitive and able to tolerate occasional inconsistencies
\cite{brew:cap}. %, Gilbert:2002:BCF:564585.564601, Abadi:2012:CTM:2360751.2360959}).
%\cite{brew:cap,Gilbert:2002:BCF:564585.564601,Abadi:2012:CTM:2360751.2360959}).
 %A large body of research deals with the problem of supporting various forms of stronger-than-eventual consistency
%	in scalable storage systems and databases.
The state machine replication paradigm achieves the strongest possible form of consistency
	by physically serializing operations \cite{lam_time}.
Lamport's Paxos protocol is a quorum-based fault-tolerant protocol for state machine replication \cite{DBLP:conf/opodis/Lamport02}.
Mencius improves upon Paxos by ensuring better scalability and higher throughput under high client load
	using a rotating coordinator scheme \cite{Mao:2008:MBE:1855741.1855767}.
  Using variations of Paxos, a number of systems \cite{journals/corr/abs-1103-2408, Bolosky:2011:PRS:1972457.1972472, Kraska:2013:MMC:2465351.2465363, Corbett:2012:SGG:2387880.2387905} claim to provide scalability and fault-tolerance. %:
%Spinnaker \cite{journals/corr/abs-1103-2408}, Gaios \cite{Bolosky:2011:PRS:1972457.1972472},
%MDCC \cite{Kraska:2013:MMC:2465351.2465363}, and Spanner \cite{Corbett:2012:SGG:2387880.2387905}.
%These systems use multiple instance of Paxos for scalability, and as a result they incorporate additional mechanisms,
%	such as two-phase commit in Spanner, to support distributed transactions.
% \par Several other papers discuss techniques for supporting transactional semantics at scale, including
%	reducing latency under low contention using ``fast'' Paxos \cite{Kraska:2013:MMC:2465351.2465363},
%	executing transactions using a combination of eventually consistent and strongly consistent operations \cite{183989,Li:2012:MGS:2387880.2387906}, % RedBlue
%	providing eventually consistent transactions \cite{Burckhardt:2012:ECT:2259248.2259252},
%	and using static analysis to prevent conflicts among transactions \cite{Zhang:2013:TCA:2517349.2522729}. % transaction chains
%Whereas Paxos-based and transactional systems aim to provide strong forms of consistency,
%	our framework is intended to simplify the tuning of systems that support simple read and write operations with
%	weak consistency, for example using sloppy quorums.
       Relatively few systems \cite{conf/wecwis/YuV00, Terry:2013:CSL:2517349.2522731, Ardekani:2014:SGC:2685048.2685077, Sivaramakrishnan:2015:DPO:2813885.2737981} provide mechanisms for fine-grained control over consistency. %, like TACT \cite{conf/wecwis/YuV00}, Pileus \cite{Terry:2013:CSL:2517349.2522731}, Tuba \cite{Ardekani:2014:SGC:2685048.2685077}, and Quela \cite{Sivaramakrishnan:2015:DPO:2813885.2737981}.
%Yu and Vahdat propose a middleware layer for tunable availability and consistency tradeoffs (TACT)
%	that uses three metrics to express consistency requirements with respect to read and write operations:
%	numerical error, order error, and staleness \cite{conf/wecwis/YuV00}.  \cite{Terry:2013:CSL:2517349.2522731, Ardekani:2014:SGC:2685048.2685077} provide automated consistency tuning under SLAs with actual trials. % However, OptCon %tunes client-side consistency
%  takes into account the consistency-latency tradeoff.
  Li et al. \cite{183989} applies static analysis for automated consistency tuning for relational databases. %, and
%  does not address quorum stores. %\cite{Sivaramakrishnan:2015:DPO:2813885.2737981} determines the weakest consistency level that satisfies the correctness condition. %It does not consider the tradeoff between staleness and latency, and cannot adapt to SLA, workload, and network state.
%\cite{lacurts2014cicada} attempts to provide a mechanism to predict the appropriate network bandwidth
%guarantees over the cloud. \cite{Winstein:2013:TEM:2486001.2486020} proposes an approach to minimize the latency
% and maximize throughput with congestion control algorithms generated from assumptions and specifications
% the network given by the by users. \cite{Ravindranath:2013:TCU:2517349.2522717} presents Timecard - a system
% that can produce consistent response time for mobile client-server applications.
%Terry et al.\ present Pileus, a transactional key-value store that supports user-specified SLAs specified as a sequence
%	of consistency and latency targets (sub-SLAs) ordered by a user-defined utility metric \cite{Terry:2013:CSL:2517349.2522731}.
%In contrast to these systems our framework can be applied on top of an existing eventually consistent data store.
%The sub-SLAs we consider are inspired by Pileus but address a narrower set of consistency targets specified as staleness bounds.

\section{Conclusions}\label{sec:conclusion}
OptCon automates client-centric consistency-latency tuning in quorum-replicated stores, based on staleness and latency thresholds in the SLAs. %, which is otherwise manually selected by skilled users.
 %OptCon provides an intelligent tradeoff between consistency and latency depending on the workload, network and system state.
        %We implement the OptCon Learning module using various Learning algorithms and do a comparative analysis of the accuracy and speed for each case (refer Table \ref{table:selection}).
         %Our comparative analysis  of the various learning techniques
%In terms of accuracy the decision tree performed best, with an area under the ROC curve of 0.9857 and Gini coefficient of 0.9714.
%The Bayes network was the runner-up with an area under the ROC curve of around 0.9 and a Gini coefficient of 0.8.
%\par The linear regression model was a partial success in terms of ANOVA analysis, with p-values below 0.01 showing that latency is strongly dependent on the input variables, especially the read/write ratio.
%For staleness we observed a strong dependence on the read/write ratio only. %, and for the retransmission rate we observed no dependence.
%Thus, latency and staleness are both amenable to prediction using linear regression. %, retransmission rate can be considered as additional parameter.
 % We compared the performance of OptCon against various fixed manually tuned client-side consistency levels. %, with varying workload using YCSB and RUBBoS benchmark suite.
   OptCon can adapt to variations in the workload, and is at least as effective as any manually chosen consistency setting %it yields 85\% success rate
  in satisfying a given SLA. %in predicting matching consistency levels under given SLAs.
   OptCon provides insights for applying machine learning to similar problems.
%In contrast, a manually chosen consistency level can optimize for either latency or staleness, but not both.
 %OptCon demonstrated adaptability to varying workload, in contrast with fixed manually chosen consistency settings.
%\par Currently, to provide the users with minimum possible staleness, we choose the strongest consistency setting in case of multiple consistency options satisfying a given subSLA. We plan to use boosting \cite{Flach:2012:MLA:2490546} in such cases to obtain the matching consistency. In future plan to work on devising a mechanism for predicting
%fallback consistency levels in case of failure of the predicted consistency settings under partition.

\renewcommand{\bibfont}{\footnotesize}
%\bibliographystyle{IEEEtran}
%\bibliography{Consistency1}
\bibliographystyle{abbrv}
\bibliography{Consistency}
% that's all folks
\end{document}